\documentclass[preprint,authoryear,12pt]{elsarticle}

 \usepackage{graphicx}
\usepackage{amssymb}
\usepackage{amsthm}
\usepackage{amsmath}
\usepackage{amsfonts}

\journal{Computer Methods in Applied Mechanics and Engineering}

\begin{document}

\begin{frontmatter}



\title{A fully nonlinear iterative solution method for self-similar
potential flows with a free boundary}


\author{A. Iafrati}

\address{INSEAN-CNR - The Italian Ship Model Basin, Rome, Italy}

\begin{abstract}
An iterative solution method for fully nonlinear boundary
value problems governing self-similar flows with a free boundary is 
presented. 
Specifically, the method is developed for application to water entry
problems, which can be studied under the assumptions of an ideal and
incompressible fluid with negligible gravity and surface tension
effects.
The approach is based on a pseudo time stepping procedure, which uses a
boundary integral equation method for the solution of the Laplace
problem governing the velocity potential at each iteration.
In order to demonstrate the flexibility and the capabilities of the
approach, several applications are presented: the classical wedge entry
problem, which is also used for a validation of the approach,
the block sliding along an inclined sea bed, the vertical water entry of a 
flat plate and the ditching of an inclined plate.
The solution procedure is also applied to cases in which the body surface is
either porous or perforated. Comparisons with numerical or experimental data 
available in literature are presented for the purpose of validation.

\end{abstract}


\end{frontmatter}

% \linenumbers

\section{Introduction}
\label{Introduction}

The water entry flow is highly nonlinear and is generally characterised by 
thin jets, as well as sharp velocity and pressure gradients. 
As a free boundary problem, the
solution is further complicated from the mathematical viewpoint by the
fact that a portion of the domain boundary, which is the free
surface, is unknown and has to be derived as a part of the solution.
At least in the early stage of the water entry, viscous effects are 
negligible and thus the fluid can be considered as ideal. Moreover, 
provided the angle between the free surface and the tangent to
the body at the contact point is greater than zero, compressible effects are
negligible \cite{korobkin88} and the fluid can be considered as
incompressible. 

Several approaches dealing with water entry problems have been developed 
within the potential flow approximation over the last twenty years,
which provide the solution in the time domain, e.g. \cite{zhao93,
battistin03, mei99, wu04, xu08, xu10} just to mention a few of them.
Time domain approaches are characterised by a high level of flexibility,
as they can be generally applied to almost arbitrary body shapes and
allow to account for the variation of the penetration velocity in time.
However, depending on the shape of the body and on the entry velocity,
the solution can be profitably written in a self-similar form by using
a set of time dependent spatial variables. In this way, the initial
boundary value problem is transformed into a boundary value problem,
e.g. \cite{semenov06, faltinsen08}.
It is worth noticing that sometime the solution formulated in terms of
time dependent variables is not exactly time independent, but it can be
approximated as such under some additional assumptions.
Problems of this kind are for instance those discussed in 
\cite{king94, iafrati04, needham08} where the solutions can be considered 
as self-similar in the limit as $t \to 0^+$.

Although avoiding the time variable reduces the complexity significantly, 
the problem remains still nonlinear as a portion of the boundary is
still unknown and the conditions to be applied over there depends on the
solution.
Several approaches have been proposed for the solution of the fully
nonlinear problem since the very first formulation by \cite{dobro69}, 
who expressed the solution of the problem in terms of a nonlinear, singular, 
integrodifferential equation.
The derivation of the solution function is rather tricky, though
\citep{zhao93}.
An approach which has some similarities with that proposed by
\cite{dobro69}, has been proposed in \cite{semenov06} where the solution is
derived in terms of two governing functions, which are the complex
velocity and the derivative of the complex potential defined in a
parameter domain. The two functions are obtained as the solution of
a system of an integral and an integro-differential equation in terms
of the velocity modulus and of the velocity angle to the free surface,
both depending on the parameter variable. This approach proved to be
rather accurate and flexible \citep{faltinsen08,semenov09}, although 
it is not clear at the moment if that approach can be easily extended to
deal with permeable conditions at the solid boundaries.

Another approach is proposed in \cite{needham08}, which generalise the
method also used in \cite{king94} and \cite{needham07}, uses a Newton
iteration method for the solution of the nonlinear boundary value problem.
The boundary conditions are not enforced in a fully nonlinear, though.
Both the conditions to be applied and the surface onto which the
conditions are applied are determined in an asymptotic manner, which is
truncated at a certain point.

In this work an iterative, fully nonlinear, solution method for 
a class of boundary value problems with free boundary is presented.
The approach is based on a pseudo-time-stepping procedure, basically 
similar to that adopted in time domain simulations of the water entry of 
arbitrary shaped bodies \citep{battistin03, iafrati03, iafrati07}.
Differently from that, the solution method exploits a modified velocity
potential which allows to significantly simplify the boundary conditions
at the free surface.
By using a boundary integral representation of the velocity
potential, a boundary integral equation is obtained by enforcing the
boundary conditions. A Dirichlet condition is applied on the
free surface, whereas a Neumann condition is enforced at the body
surface.
In discrete form, the boundary is discretized by straight line panels
and piecewise constant distributions of the velocity potential and of its
normal derivative are assumed.

As already said, water entry flows are generally characterised by thin 
jets developing along the body surface or sprays detaching from the edges 
of finite size bodies.
An accurate description of the solution inside such thin layers requires
highly refined discretizations, with panel size of the order of the jet
thickness. A significant reduction of the computational effort can be
achieved by cutting the thin jet off the computational domain without
affecting the accuracy of the solution substantially.
However, there are some circumstances in which it is important to extract
some additional information about the solution inside the jet, e.g. the 
jet length or the velocity of the jet tip. 
For those cases, a shallow water model has been developed which allows
to compute the solution inside the thinnest part of the jet in an
accurate and efficient way by exploiting the hyperbolic structure of the
equations. A space marching procedure is adopted, which is started at the 
root of the jet by matching the solution provided by the boundary integral 
representation.

The solution method is here presented and applied to several water entry
problems. It is worth noticing that the method has been adopted in the
past to study several problems, but was never presented in a unified
manner. For this reason, in addition to some brand new results, some 
results of applications of the method which already appeared in 
conference proceedings or published paper are here briefly reviewed. 
The review part is not aimed at discussing the physical aspects but mainly 
at showing, with a unified notation, how the governing equations change when 
the model is applied to different contexts.
Solutions are presented both for self-similar problems and for problems
which are self-similar in the small time limit. Applications are also
presented for bodies with porous or perforated surfaces where the
the boundary condition at the solid boundary depends on the local
pressure.

The mathematical formulation and the boundary value problem are derived
in section 2 for the self-similar wedge entry problem with constant
velocity.
The disrete method is illustrated in section 3, together with a
discussion over the shallow water model adopted for the thin jet layer. 
The applications of the proposed approach to the different problems 
are presented in section 4 along with a discussion about the changes in
the governing equations.

\section{Mathematical formulation}
\label{MathForm}

The mathematical formulation of the problem is here derived referring to
the water entry of a two-dimensional wedge,
under the assumption of an ideal and incompressible fluid with
gravity and surface tension effects also neglected.
The wedge, which has a deadrise angle $\gamma$, touches the free surface
at $t=0$ and penetrates the water at a constant speed $V$.
Under the above assumptions, the flow can be expressed in terms of a velocity
potential $\phi(x,y,t)$ which satisfies the initial-boundary value problem
\begin{eqnarray}
\nabla^2 \phi & = & 0 \qquad \qquad \Omega(t)  \\
\label{simcon}
\frac{\partial \phi}{\partial n} & = & 0  \qquad \qquad x=0 \\
\frac{\partial \phi}{\partial n}& = & V \cos \gamma \qquad y=-Vt+x \tan
\gamma \\
\label{dbcfis}
\frac{\partial \phi}{\partial t} + \frac{1}{2} {\boldsymbol u}^2 &
= & 0 \qquad \qquad H(x,y,t) = 0 \\
\label{kbcfis}
\frac{D H}{D t} & = & 0 \qquad \qquad H(x,y,t) = 0 \\
H(x,y,0) & = & 0  \qquad \qquad y = 0 \\
\phi(x,y,0) & = & 0 \qquad \qquad y = 0 \\
\phi(x,y,t) & \to & 0 \qquad \qquad (x^2+y^2) \to \infty
\end{eqnarray}
where $\Omega(t)$ is the fluid domain, $H(x,y,t)=0$ is the equation of the
free surface, $\boldsymbol n$ is the normal to the fluid domain oriented
inwards and $x,y$ represent the horizontal and vertical coordinates,
respectively. Equation (\ref{simcon}) represents the symmetry condition
about the $y$ axis. In equation (\ref{dbcfis}) $\boldsymbol u = \nabla
\phi$ is the fluid velocity.
It is worth remarking that the free surface shape, i.e.
the function $H(x,y,t)$, is unknown and has to be determined as a part of
the solution. This is achieved by satisfying equations (\ref{dbcfis})
and (\ref{kbcfis}) which represent the dynamic and kinematic boundary
conditions at the free surface, respectively.

By introducing the new set of variables:
\begin{equation}
\label{selfvar}
\xi = \frac{x}{Vt} \qquad
\eta = \frac{y}{Vt} \qquad
\varphi = \frac{\phi}{V^2t} \;\;,
\end{equation}
the initial-boundary value problem can be recast into a
self-similar problem, which is expressed by the following set of
equations:
\begin{eqnarray}
\nabla^2 \varphi & = & 0 \qquad \qquad \varOmega  \\
\varphi_\xi & = & 0 \qquad  \qquad\xi=0 \\
\label{bbcss}
\varphi_\nu & = & \cos \gamma \qquad \eta =-1 + \xi \tan \gamma \\
\label{dbcss}
\varphi - (\xi \varphi_\xi + \eta \varphi_\eta) + \frac{1}{2} \left(
\varphi_\xi^2 + \varphi_\eta^2 \right) &=& 0  \qquad\qquad h(\xi,\eta) = 0 \\
\label{kbcss}
- (\xi h_\xi + \eta h_\eta) + (h_\xi \varphi_\xi + h_\eta \varphi_\eta)
&=& 0 \qquad \qquad h(\xi,\eta) = 0  \\
\label{ffb}
\varphi & \to & 0 \qquad \qquad \xi^2 + \eta^2 \to \infty
\end{eqnarray}
where $\varOmega$ is fluid domain, $h(\xi,\eta)=0$ is the equation of
the free surface and $\nu$ is the unit normal to the boundary, which is
oriented inward the fluid domain.

\begin{figure}
\centerline{\hbox{
\includegraphics[height=60mm]{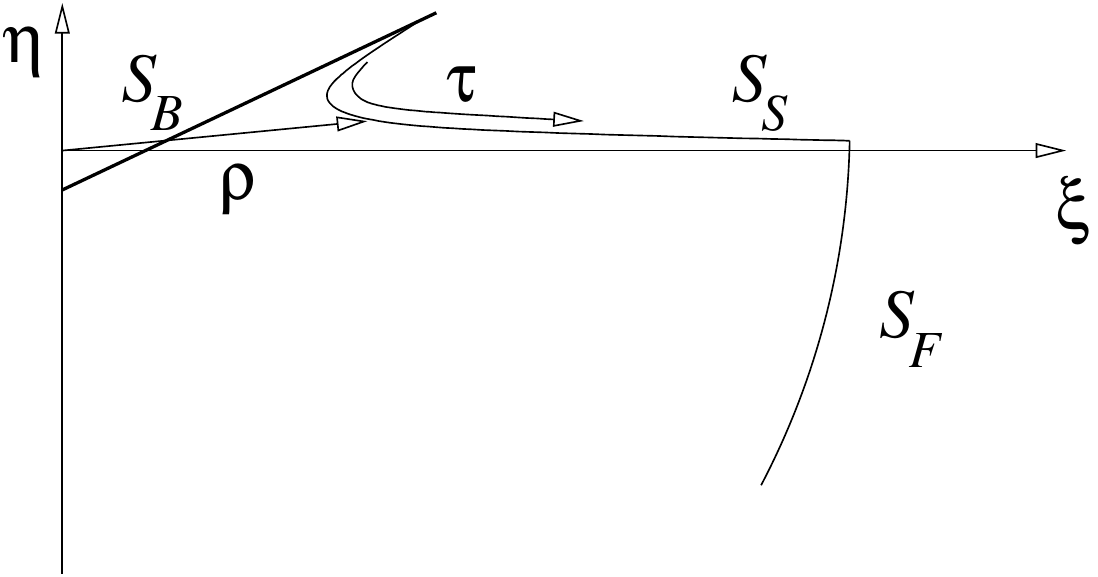}
}}
\caption{Sketch of the computational domain and of the notation adopted.\label{sck}}
\end{figure}

Despite the much simpler form, the boundary value problem
governing the self-similar solution is still rather challenging as the
free surface shape is unknown and nonlinear boundary conditions have to
be enforced on it.
The free surface boundary conditions can be significantly simplified by
introducing a modified potential $S(\xi,\eta)$ which is defined as
\begin{equation}
\label{defS}
S(\xi,\eta) = \varphi(\xi,\eta) - \frac{1}{2} \rho^2
\end{equation}
where $\rho = \sqrt{\xi^2+\eta^2}$.
By substituting equation (\ref{defS}) into the kinematic condition
(\ref{kbcss}), it follows that $\nabla S \cdot \nabla h = 0$, which is
\begin{equation}
\label{kbcssm}
S_\nu = 0
\end{equation}
on the free surface $h=0$.
Similarly, by substituting equation (\ref{defS}) into the dynamic
condition (\ref{dbcss}), it is obtained that
\[
S + \frac{1}{2} \left( S_\xi^2 + S_\eta^2 \right) = 0
\]
on the free surface. By using equation (\ref{kbcssm}), the above
equation becomes
\begin{equation}
S + \frac{1}{2} S_\tau^2 = 0 \qquad \Rightarrow \qquad S_\tau = \pm
\sqrt{-2 S} \;\;,
\end{equation}
where $\tau$ is the arclength measured along the free surface
(Fig.~\ref{sck}).

By combining all equations together, we arrive at the new boundary value
problem
\begin{eqnarray}
\label{lapss}
\nabla^2 \varphi & = & 0 \qquad \qquad \qquad \varOmega  \\
\varphi_\xi & = & 0 \qquad \qquad  \qquad\xi=0 \\
\label{bbcf}
\varphi_\nu & = & \cos \gamma \qquad \qquad \qquad
\eta =-1 + \xi \tan \gamma \\
\label{sdef}
\varphi &=& S + \frac{1}{2} \rho^2  \qquad\qquad h(\xi,\eta) = 0 \\
\label{dbcssf}
S_\tau &=& \pm \sqrt{-2 S}  \qquad\qquad h(\xi,\eta) = 0 \\
\label{kbcssf}
S_\nu &=& 0  \qquad \qquad\qquad h(\xi,\eta) = 0 \\
\label{sff}
S & \to & - \frac{1}{2} \rho^2 \qquad \qquad \rho \to \infty
\end{eqnarray}
solution of which is derived numerically via a pseudo-time stepping procedure
discussed in detail in the next section.

\section{Iterative Solution Method}
\label{IterMeth}

The solution of the boundary value problem (\ref {lapss})-(\ref
{kbcssf}) is obtained by a pseudo-time stepping procedure similar to
that adopted for the solution of the water entry problem in time domain 
\citep{battistin03,battistin04}.
The procedure is based on an Eulerian step, in which the boundary
value problem for the velocity potential is solved,
and a Lagrangian step, in which the free surface
position is moved in a pseudo-time stepping fashion and the 
velocity potential on it is updated.

\subsection{Eulerian substep}
Starting from a given free surface shape with a corresponding distribution
of the velocity potential on it, the velocity potential at any point
$\boldsymbol x_P = (\xi_P,\eta_P) \in \:\: \varOmega$ is written in the
form of boundary integral representation 
\begin{equation}  \label{bem}
\varphi( \boldsymbol x_P) = \int_{S_S \cup S_B \cup S_\infty}
\left(\frac{\partial \varphi(\boldsymbol x_Q)}
{\partial \nu_Q} G(\boldsymbol x_P-\boldsymbol x_Q) - 
\varphi(\boldsymbol x_Q) \frac{\partial G (\boldsymbol x_P-
\boldsymbol x_Q)}{\partial \nu_Q} \right) \mbox{d}S_Q
\; ,
\end{equation}
where $\boldsymbol x_Q = (\xi_Q,\eta_Q) \in \:\: \partial \varOmega$. 
In the integrals, $S_B$ and $S_S$ denote the body contour and free surface,
respectively, $S_{\infty}$ the boundary at infinity and 
$G(\boldsymbol x)=\log\left(|\boldsymbol x|\right)/2 \pi$ 
the free space Green's function of the Laplace operator in
two-dimensions.

The velocity potential along the free surface is assigned by
equations (\ref{sdef}) and (\ref{dbcssf}), whereas its normal derivative
on the body contour is given by equation (\ref{bbcf}). In order to
derive the velocity potential on the body and the normal derivative at
the free surface the limit of equation (\ref{bem}) is taken as $\boldsymbol
x_P \to \partial \varOmega$.
Under suitable assumptions of regularity of the fluid boundary,
for each $\boldsymbol x_P\in \:\: \partial \varOmega$ it is obtained
\begin{equation}  \label{bbem}
\frac{1}{2}\varphi( \boldsymbol x_P) = \int_{S_S \cup S_B \cup S_\infty}
\left( 
\frac{\partial \varphi(\boldsymbol x_Q)}
{\partial \nu_Q} G(\boldsymbol x_P-\boldsymbol x_Q) -
\varphi(\boldsymbol x_Q) \frac{\partial G (\boldsymbol x_P-
\boldsymbol x_Q)}{\partial \nu_Q} \right) \mbox{d}S_Q
\: .
\end{equation}
The above integral equation with mixed Neumann and Dirichlet boundary 
conditions, is solved numerically by discretizing the domain boundary
into straight line segments along which a piecewise constant distribution of
the velocity potential and of its normal derivative are assumed. 
Hence, in discrete form, it is obtained
\begin{equation}
\label{discre1}
a_i \varphi_i + \sum_{j=1}^{N_B} \varphi_j d_{ij} -
\sum_{j=N_B+1}^{N_B+N_S} \varphi_{\nu,j} g_{ij} = e_i
\varphi_i - \sum_{j=N_B}^{N_B+N_S} \varphi_j d_{ij} +
\sum_{j=1}^{N_B} \varphi_{\nu,j} g_{ij} \;\;,
\end{equation}
where $N_B$ and $N_S$ indicate the number of elements on the body and on
the free surface, respectively, with $(a_i,e_i)=(1/2,0)$ if $i \in (1, N_B)$
and $(a_i,e_i)=(0,-1/2)$ if $i \in (N_B, N_B+N_S)$.
In (\ref{discre1}) $g_{ij}$ and $d_{ij}$ denote the influence
coefficients of the segment $j$ on the midpoint of the segment $i$
related to $G$ and $G_\nu$, respectively.
It is worth noticing that the when $\boldsymbol x_P$ lies onto one of
the segments representing the free surface, the integral of the influence 
coefficient $d_{ij}$ is evaluated as the Cauchy principal part.
The symmetry condition about the $\xi=0$ axis is enforced by
accounting for the image when computing the influence coefficients.

The size of the panels adopted for the discretization is refined during
the iterative process in order to achieve a satisfactory accuracy in the
higly curved region about the jet root. Far from the jet root
region, the panel size grows with a factor which is usually 1.05.

The linear system (\ref{discre1}) is valid provided the computational
domain is so wide that condition (\ref{ffb}) is satisfied at the desired
accuracy at the far field.
A significant reduction of the size of the domain can be achieved by
approximating the far field behaviour with a dipole solution. When such
an expedient is adopted, a far field boundary $S_F$ is introduced at a
short distance from the origin, along which the velocity potential is 
assigned as $C_D \varphi_D$ where $\varphi_D$ is the dipole solution
\[
\varphi_D = \frac{\eta}{\xi^2 + \eta^2} \;\;.
\]
Along the boundary $S_F$, the normal derivative is derived from the 
solution of the boundary integral equation. 
Let $N_F$ denote the number of elements located on the far field boundary, 
the discrete system of equations becomes
\begin{eqnarray}
\nonumber
a_i \varphi_i + \sum_{j=1}^{N_B} \varphi_j d_{ij} -
\sum_{j=N_B+1}^{N_B+N_S+N_F} \varphi_{\nu,j} g_{ij} + C_D
\sum_{j=N_B+N_S+1}^{N_F} \varphi_{D,j} d_{ij} = e_i
\varphi_i - &~& \\
\label{discre2}
\sum_{j=N_B+1}^{N_B+N_S+N_F} \varphi_j d_{ij} +
\sum_{j=1}^{N_B} \varphi_{\nu,j} g_{ij} \;\;, &~&
\end{eqnarray}
where $i \in (1, N_B+N_S+N_F)$.
An additional equation is needed to derive the constant of the dipole
$C_D$
together with the solution of the boundary value problem.
There is no unique solution to assign such additional condition. Here
this additional equation is obtained by enforcing the condition that
the total flux across the far field boundary has to equal that associated 
to the dipole solution, which in discrete form reads \citep{battistin04}
\begin{equation}
\label{ffdi}
- \sum_{j=N_B+N_S+1}^{N_B+N_S+N_F} \varphi_{\nu,j} \Delta s_j +
C_D \sum_{j=N_B+N_S+1}^{N_B+N_S+N_F} \varphi_{D\nu,j} \Delta s_j
= 0 \;\;.
\end{equation}

The solution of the linear system composed by equations (\ref{discre2})
and (\ref{ffdi}), provides the velocity potential on the body contour
and its normal derivative on the free surface.
Hence, the tangential and normal derivatives of the modified velocity
potential can be computed as
\begin{equation}
S_\tau = \varphi_\tau - \rho \rho_\tau  \qquad , \qquad
S_\nu = \varphi_\nu - \rho \rho_\nu
\end{equation}
allowing to check if the kinematic condition on the free
surface (\ref{kbcssf}) is satisfied. Unless the condition is satisfied
at the desired accuracy, the free surface shape and the distribution of 
the velocity potential on it are updated and a new iteration is made.

\subsection{Update of free surface shape and velocity potential}

The solution of the boundary value problem makes available the normal 
and tangential derivatives of $S$ at the free surface. A new
guess for the free surface shape is obtained by displacing the free surface
with the pseudo velocity field $\nabla S $, which is
\begin{equation}
\label{eqadv}
\frac{D \boldsymbol x}{D t} = \nabla S \;\;.
\end{equation}
Equation (\ref{eqadv}) is integrated in time (actually, it would be more
correct to say pseudo time) by a second order Runge Kutta scheme.
The time interval is chosen so that the displacement of the centroid in
the step is always smaller than one fourth of the corresponding panel size.
Once the new shape is available, the modified velocity potential on it
is initialized and the velocity potential is derived from 
equation (\ref{sdef}).
At the intersection of the free surface with the far field boundary the
velocity potential is provided by the dipole solution, by using the
constant of the dipole obtained from the solution of the boundary value
problem at the previous iteration. 
The value of the velocity potential is used to compute the
corresponding modified velocity potential from equation (\ref{sdef})
and then the
dynamic boundary condition (\ref{dbcssf}) is integrated along
the free surface moving back towards the intersection with the body contour,
thus providing the values of $S$ and $\varphi$ on the new guess.

For the wedge entry problem, at the far field, the modified velocity
potential behaves as $S \to -\rho^2/2$ and thus, $S_\tau \simeq - \rho$
as $\rho \to \infty$ (Fig.~\ref{sck}). In this case the boundary
condition can be easily integrated along the free surface, in the form
\begin{equation}
\label{sana}
S = -\frac{1}{2} (\tau + C)^2 \;\;.
\end{equation}
By assuming that the free surface forms a finite angle with the body
contour, the boundary conditions (\ref{bbcf}) and (\ref{kbcssf}) can be
both satisfied at the intersection point only if $S_\tau=0$ and thus
$S=0$. From equation (\ref{sana}) it follows that the conditions are
satisfied if $C=0$ and $\tau$ is taken with origin at the intersection
point, i.e.
\begin{equation}
\label{sana2}
S = -\frac{1}{2} \tau^2 \;\;.
\end{equation}
Although equation (\ref{sana2}) holds for the final solution, the
conditions are not satisfied for an intermediate solution.
In this case, for the new free surface shape the curvilinear
abscissa $\tau$ is initialized starting from the intersection with
the body contour and the constant $C$ is chosen to match the value of
the modified velocity potential at the intersection with the far field
boundary.
Once the distribution of the modified velocity potential on the free
surface is updated, the velocity potential is derived from equation
(\ref{sdef}), and the boundary value problem can be solved.

\begin{figure}
\centerline{\hbox{
\includegraphics[height=25mm]{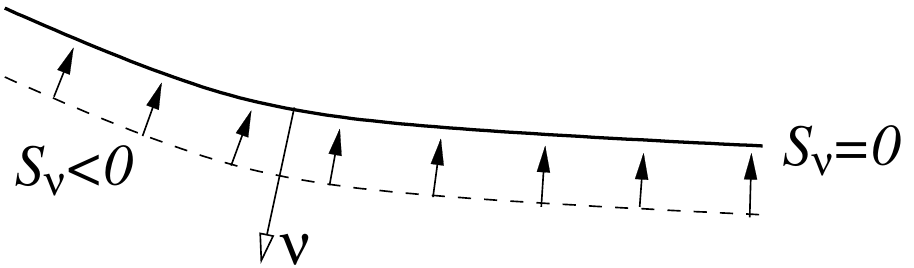}
}}
\caption{Orientation of the normal velocity field on the free surface guess.\label{sck3}}
\end{figure}

Additional considerations are deserved by the choice of $\nabla S$ as
pseudo velocity field.
It is easy to see that with such a choice, once the final solution
has been reached and $S_\nu $ approaches zero all along the free surface,
the displacements of the centroids are everywhere tangential to the
free surface, thus leaving the free surface shape unchanged.
However, this is not enough to explain why the use of $S_\nu$ as normal
velocity drives the free surface towards the solution of the problem.
By differentiating twice in $\tau$ equation (\ref{sana}) we have that
$S_{\tau \tau} = -1$.
From equation (\ref{defS}), it is $S_{\nu \nu} + S{\tau
\tau} = -2$ and then on the free surface
\[
S_{\nu \nu} = -2 - S_{\tau \tau} = -1 \;\;.
\]
According to the kinematic condition (\ref{kbcssf}) $S_\nu = 0$ on the
free surface, so that if $S_{\nu\nu} < 0 $ we have $S_\nu < 0 $ in the
water domain, which implies that using $S_\nu$ as normal velocity drives
the free surface towards the solution (Fig.~\ref{sck3}).

\subsection{Jet modelling}

Water entry flows often generates thin sprays along the body contours. 
An accurate description of such thin layer by a boundary integral
representation requires highly refined discretizations, with panel
dimensions comparable to the local thickness.
Beside increasing the size of the linear system to be solved, the small
panel lenghts in combination to the high velocity characterizing the
jet region yields a significant reduction of the time step making a
detailed description of the solution highly expensive from the
computational standpoint.

Despite the effort required for its accurate description, the jet does
not contribute significantly to the hydrodynamic loads acting on the body,
which is doubtless the most interesting quantity to be evaluated in a
water entry problem.
Indeed, due to the small thickness, the pressure inside the spray is
essentially constant and equal to the value it takes at the free surface,
which is $p = 0$. This is the reason why rather acceptable estimates of
the pressure distribution and total hydrodynamic loads can be obtained 
by cutting
off the thinnest part of the jet, provided a suitable boundary condition
is applied at the truncation. In the context of time domain solutions,
the cut of the jet was exploited for instance in \cite{battistin03}
in which the jet is cut at the point where the angle between the
free surface and the body contour drops below a threshold value.

\begin{figure}
\centerline{\hbox{
\includegraphics[height=25mm]{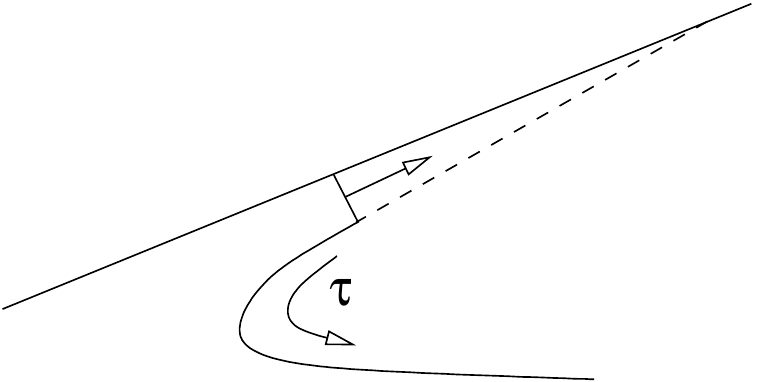}
\includegraphics[height=25mm]{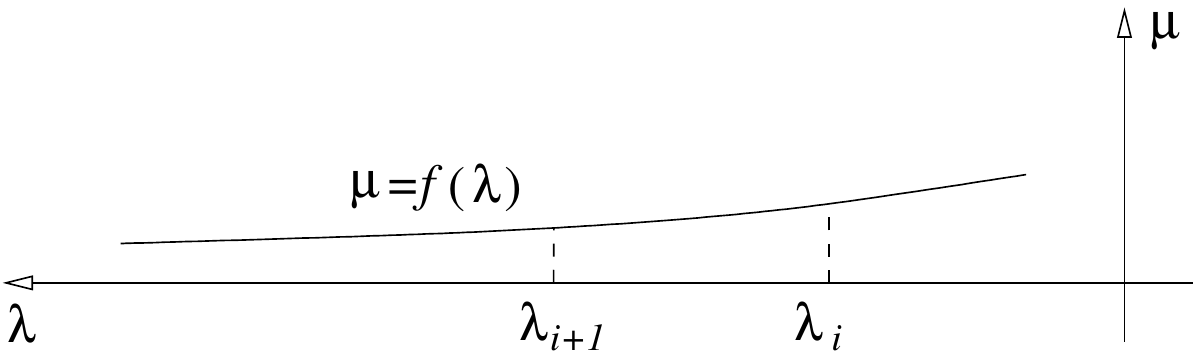}
}}
\caption{Model adopted for the thin jet and local coordinate system for
the shallow water model.  \label{sck4}}
\end{figure}

A similar cut can be adopted in the context self-similar problems
discussed here. The truncated part is accounted for by assuming that the 
normal velocity at the truncation of the jet equals the projection on
the normal direction of the velocity at the free surface. 
From the sketch provided in Fig.~\ref{sck4}, $\tau^*$, which denotes
the value of the arclength at the intersection between the free surface 
and the jet truncation, is different from zero and thus, from equation 
(\ref{sana2}) it follows 
\begin{equation}
\label{veltantappo}
S_\tau(\tau^*) = -\tau^* \;\;.
\end{equation}
The value $\tau^*$ is derived by initializing the arclength
on the free surface as $\tau=\tilde \tau + \tau^*$ and
matching the velocity potential at the far field with the asymptotic
behaviour. If the computational domain is large enough, the matching is 
established in terms of the tangential velocity and thus, from equation 
(\ref{sff}), we simply get that
\[
\tau^* = \rho - \tilde \tau \;\;.
\]
When the dipole solution $C_D \varphi_D$ is used to approximate the 
far field solution, $\tau^*$ is determined by matching the velocity 
potential and then, by equations (\ref{sdef}) and (\ref{sana2}), we get
\[
\tau^* = (\rho^2-2 C_D \varphi_D)^{1/2}- \tilde \tau \;\;.
\]
The tangential velocity at the intersection of the free surface with the 
jet truncation line can be computed from
\begin{equation}
\varphi_\tau = S_\tau + \rho \rho_\tau \;\;,
\end{equation}
where $S_\tau$ comes from equation (\ref{veltantappo}).
Hence, the normal velocity to be used as boundary condition at the 
jet truncation is derived as
\begin{equation}
\varphi_\nu = \varphi_\tau \cos \beta \;\;,
\end{equation}
where $\beta$ is angle formed by body contour with the tangent at the
free surface taken at the intersection with jet truncation line.

Although very efficient and reasonably accurate, such simplified models 
do not allow to derive any information in terms of wetted 
surface and jet speed.
An alternative approach, which provides all the information in a rather
efficient way, exploits the shallowness of the jet layer.
In order to explain the model, it is useful to consider the governing 
equations in a local frame of reference with $\lambda$ denoting the coordinate
along the body and $\mu = f(\lambda)$ the local thickness (Fig.~\ref{sck4}).
The kinematic boundary condition (\ref{kbcssf}) becomes
\begin{equation}
\label{kbcsha}
S_\mu = S_\lambda f_\lambda \qquad \mbox{on } \; \mu = f(\lambda) \;\;.
\end{equation}

From the definition (\ref{defS}), it follows that $\nabla^2 S=-2$.
Integration of the above equation across the jet thickness, i.e.
along a $\lambda = const$ line, provides 
\begin{equation}
\int_0^{f(\lambda)} S_{\lambda \lambda} (\lambda, \mu) \mbox{d} \mu +
\left[ S_\mu(\lambda, \mu) \right]_{\mu=0}^{\mu=f(\lambda)} = -2
f(\lambda) \;\;.
\end{equation}
By exploting the body boundary conditions on the body surface, we
finally get \citep{korobkin06}
\begin{equation}
\label{eqsh}
\frac{\mbox{d}}{\mbox{d}\lambda} \int_0^{f(\lambda)} S_{\lambda}
(\lambda, \mu) \mbox{d} \mu = -2 f(\lambda) \;\;,
\end{equation}
which can be further simplified by neglecting the variations of the
modified velocity potential across the jet layer, thus arriving 
to 
\begin{equation}
\label{eqstilde}
\left[ \tilde S_\lambda(\lambda) f(\lambda) \right]_\lambda + 2
f(\lambda) = 0  \;\;,
\end{equation}
where $\tilde S(\lambda) = S\left(\lambda, f(\lambda)\right) \simeq 
S(\lambda,\mu)$.
In terms of $\tilde S(\lambda)$ the kinematic condition (\ref{kbcsha})
reads
\begin{equation}
\label{kbcstilde}
\tilde S_\lambda = - \frac{S_\tau}{\sqrt{1+f_\lambda^2}}
\end{equation}

Equations (\ref{eqstilde}) and (\ref{kbcstilde}), together with the
dynamic boundary condition (\ref{sana2}) can be used to build an
iterative, space marching procedure which, at the jet root, matches the
solution provided by the boundary integral formulation in the bulk of the
fluid.
The most relevant point of the procedure are discussed here, whereas 
a more detailed description can be found in \cite{korobkin06}.

By using a finite difference discretization of equation
(\ref{eqstilde}), the following equation for the jet thickness is obtained
%%
%In discrete form, equation (\ref{eqstilde}) can be written as
%\begin{equation}
%\label{discresha}
%\frac{\tilde S_\lambda(i+1) f(i+1) - \tilde S_\lambda(i) f(i) }{\Delta
%\lambda} + (f(i+1)+f(i))  = 0 \;\;,
%\end{equation}
%where $\tilde S_\lambda(i+1)$ and $f(i+1)$ are both unknown and are
%derived altogether by an iterative solution of equation
%(\ref{discresha}) and (\ref{kbcstilde}). The jet thickness $f(i+1)$ is
%derived from equation (\ref{discresha}) as
%%
\begin{equation}
\label{fkip}
f^k(i+1) = \omega f^{k-1}(i+1) - (1-\omega) f(i) \left[ \frac{
\Delta \lambda - \tilde S_\lambda (i) }{\Delta \lambda + \tilde
S^{k-1}_\lambda (i+1)  } \right] \;\;.
\end{equation}
As $\tilde S_\lambda(i+1)$ and $f(i+1)$ are both unknown, the solution
is derived by subiterations. In equation (\ref{fkip}) $k$ is the 
subiteration number and $\omega$ is a relaxation parameter, which is 
usually taken as 0.9.
Once the new estimate of the local thickness $f^k(i+1)$ is available, it
is used in the kinematic condition (\ref{kbcstilde}) to evaluate
$S^{k}_\lambda (i+1) $ as
\begin{equation}
\label{slam}
S^{k}_\lambda (i+1) = - \frac{S^k_\tau(i+1)}{\sqrt{1+\left[
f_\lambda^k(i+1)\right]^2 }} \;\;,
\end{equation}
where the derivative of the thickness is evaluated in discrete form as
\begin{equation}
f_\lambda^k(i+1) = \frac{f^k(i+1)-f(i)}{\Delta \lambda} \;\;.
\end{equation}
In equation (\ref{slam}) the term $S^k_\tau(i+1)$ is estimated by
exploiting equation
(\ref{sana2}) which provides $S_\tau(i) = -\tau(i)$, and thus
\begin{equation}
\label{stauk}
S_\tau^k (i+1) = S_\tau(i) + \sqrt{ \Delta \lambda^2 + (f^k(i+1)-f(i))^2}
\;\;.
\end{equation}
The system of equations (\ref{fkip})-(\ref{stauk}) is solved by
subiterations, which use $f(i)$ and $\tilde S_\lambda(i)$ as first guess 
values.
All quantities at $i=1$ are derived from the boundary integral representation,
thus ensuring the continuity of the solution.
The spatial step $\Delta \lambda$ is assumed equal to half of the size of 
the first free surface panel attached to the jet region. 
Such a choice turns to be useful for the computation of the pressure, as 
discussed in the next section.  The space marching
procedure is advanced until reaching the condition $|S_\lambda(i+1) | < 
\Delta \lambda$, which implies
that the distance to the intersection with the body contour is smaller than
$\Delta \lambda$.

\subsection{Pressure distribution}

The pressure field on the body can be derived from the distribution 
of the velocity potential.
Let $\varrho$ denote the fluid density, the equation of the local
pressure
\[
p = - \varrho \left\{ \frac{\partial \phi}{\partial t} - \frac{1}{2}
\left| \boldsymbol u \right|^2 \right\} \;\;,
\]
is written in terms of the self-similar variables (\ref{selfvar}) thus 
leading to
\begin{equation}
\label{pressure}
\psi(\xi,\eta) = - \varphi + (\varphi_\xi \xi + \varphi_\eta
\eta) - \frac{1}{2} (\varphi_\xi^2 + \varphi_\eta^2 ) \;\;,
\end{equation}
where $\psi = p/(\varrho V^2)$ is the nondimensional pressure.

When the shallow water model is activated, due to the assumptions, the
pressure in the modelled part of the jet is constant and equal to the value
it takes at the free surface, i.e zero. As a consequence, a sharp drop
of the pressure would occur across the separation line between
the bulk of the fluid and the modelled part of the jet.
In order to avoid such artificial discontinuity, the velocity potential
along the body is recomputed by using the boundary integral
representation of the velocity potential for the whole region containing 
both the bulk of the fluid and the part of the jet modelled by the shallow 
water model.
As discussed in the previous section, the spatial step in the space marching
procedure of the shallow water model has been chosen as half of the size
of the free surface panel adjacent to the modelled part of the jet.
This allows a straightforward derivation of the discretization to be
used in the boundary integral representation. Two adjacent panels in the
shallow water region are used to define a single panel in the discrete boundary
integral representation (\ref{discre2}). This panel has the velocity
potential associated at the mid node connecting the two shallow water elements
which constitutes the panel.
On the basis of the above considerations, if $N_{SW}$ is the number of
steps in the space marching procedure, the inclusion of the modelled
part of the jet in the boundary integral representation corresponds to
$N_{SW}/2$ panels on the body contour and $N_{SW}/2$ panel on the free
surface, with a total of $N_{SW}$ additional equations in
(\ref{discre2}). 
As usual, a Neumann boundary condition is applied to the panels lying 
along the body contour, and a Dirichlet condition is applied to
the panels lying on the free surface.
It is shown in the following that including the shallow water solution
in the boundary integral representation results in a much smoother
pressure distribution about the root of the jet, whereas the remaining
part is essentially unchanged.

\subsection{Porous and perforated contours}

The solution procedure is also applicable to problems in which the 
solid boundary is permeable, provided the boundary condition can be 
formulated as a function of the pressure.
In these case, of course, an accurate prediction of the pressure
distribution is mandatory.

Possible examples are represented by porous or perforated surfaces.
In a porous surface the penetration velocity depends on a balance
between the viscous losses through the surface and the pressure jump, so
that, if the $\boldsymbol V \cdot \boldsymbol n $ is the normal velocity
of the contour, the actual boundary condition is 
\begin{equation}
\label{porous}
\frac{\partial \phi}{\partial n}  = \boldsymbol V \cdot
\boldsymbol n - \alpha_0 p \;\;,
\end{equation}
where $\alpha_0$ is a coefficient that depends on the characteristics of
the porous layer \citep{iafrati05b}.
In a perforated surface, the flow through the surface is governed by a
balance between the inertial terms and the pressure jump. In this case
the condition is usually presented in the form \cite{molin01}
\begin{equation}
\label{perforated}
\frac{\partial \phi}{\partial n}  = \boldsymbol V \cdot
\boldsymbol n - \chi \sqrt{p/\varrho} \;\;, \qquad \chi^2 = \frac{2 
\sigma \kappa^2}{1- \kappa}
\end{equation}
where $\sigma$ is a dicharge coefficient which is about $0.5$ and
$\kappa$ is the perforation ratio, i.e. the ratio between the area of
the holes and the total area.

In terms of the self-similar variables, both cases are still described
by the system of equations (\ref{selfvar})-(\ref{ffb}) but for the
body boundary condition (\ref{bbcss}) which is rewritten as
\begin{equation}
\label{ffbpp}
\varphi_\nu = \cos \gamma - f(\psi) \qquad \eta = -1 + \xi t \tan \gamma
\;\;,
\end{equation}
where $f(\psi) = \alpha_0 \psi$ and 
$f(\psi) = \chi \sqrt{\psi} $ for porous and perforated surfaces, 
respectively.

Examples of water entry flows of porous or perforated bodies are
presented in the next section. The important changes operated on the
solution by different levels of permeability are clearly highlighted.

\section{Applications}

\subsection{Wedge entry problem}

\begin{figure}
\centerline{\hbox{
\includegraphics[width=80mm]{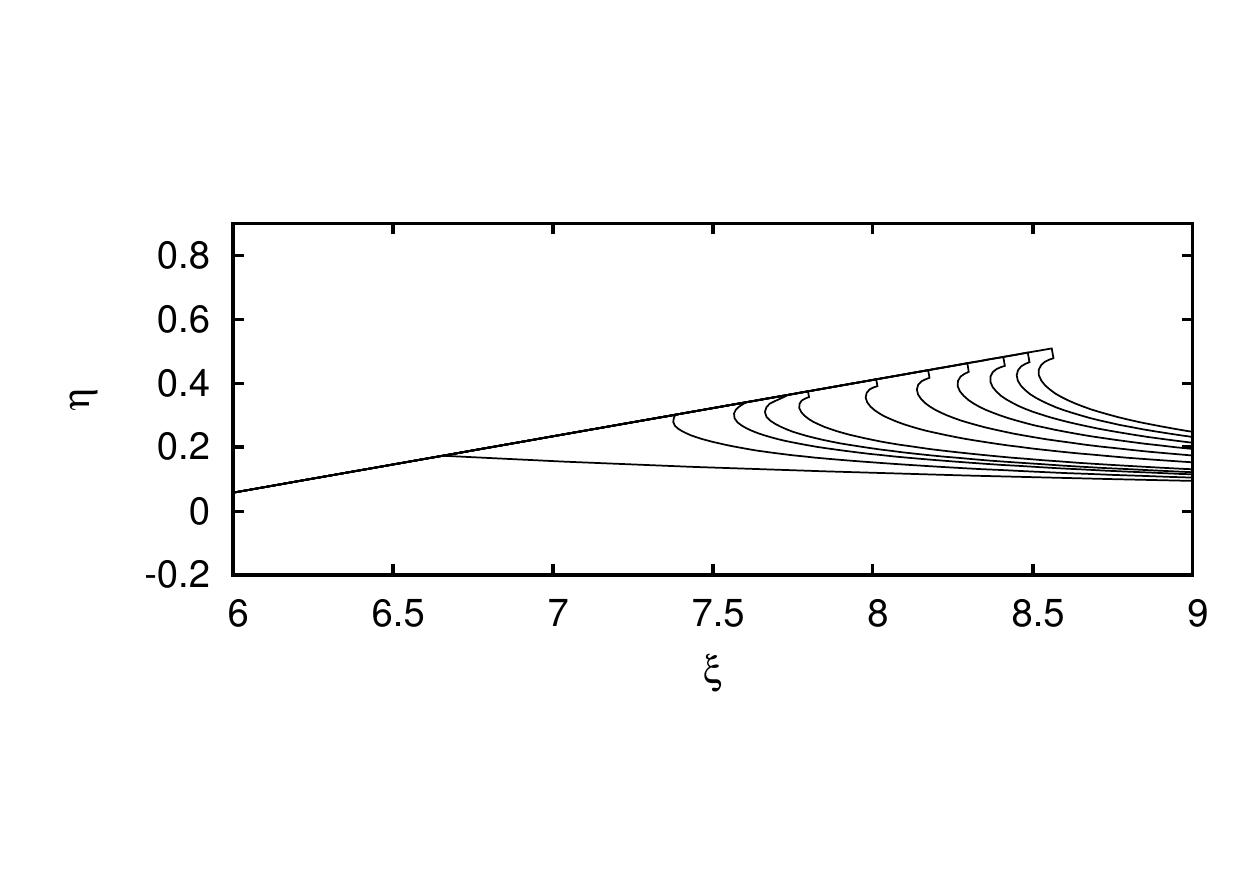}
\includegraphics[width=80mm]{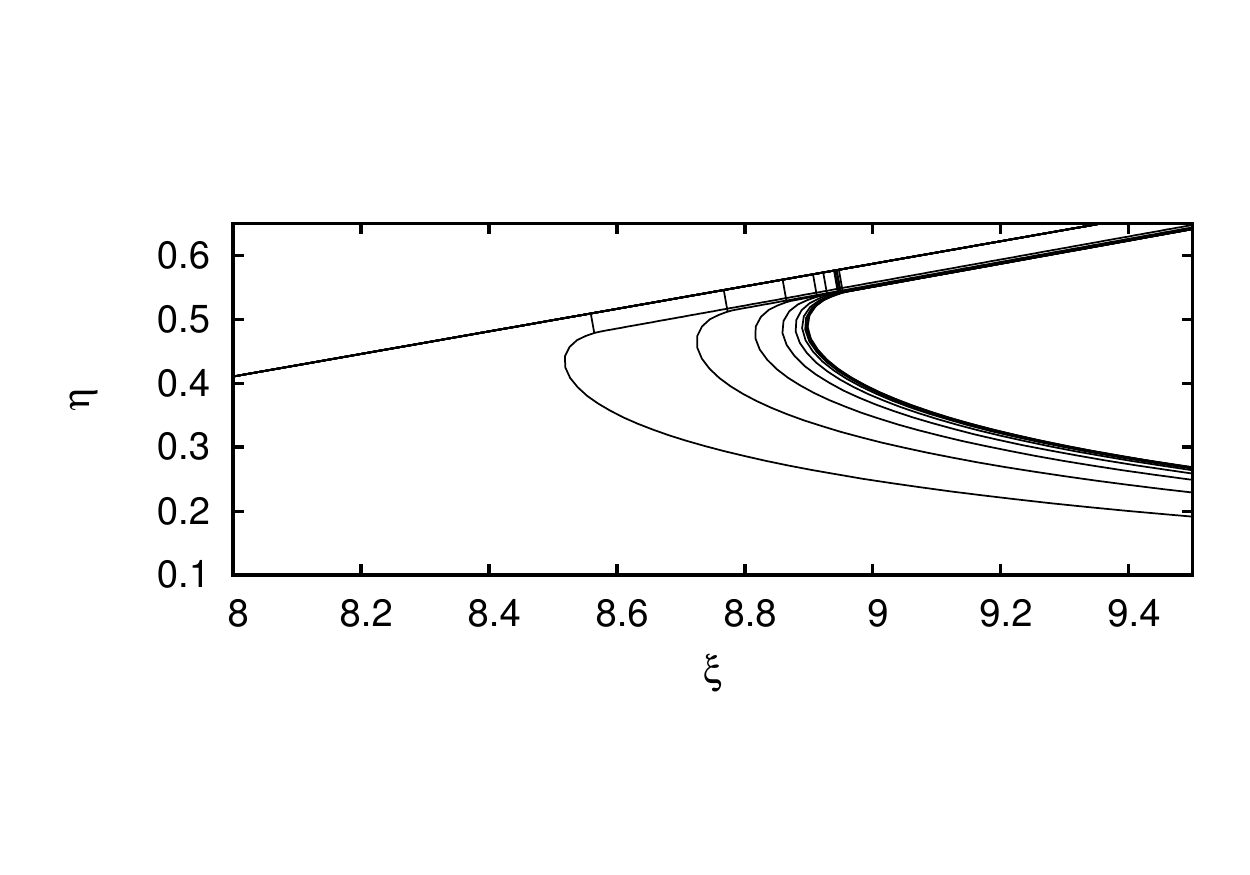}
}}
\caption{Convergence of the iterative process for a wedge with 10
degrees deadrise angle. The left picture display
the early stage of the process, starting from the initial configuration,
till the formation of the thin jet layer (not shown) where the shallow
water model is adopted. On the right picture, the convergence process
about the root of the jet is shown.\label{conv_10}}
\end{figure}

As a first application, the computational method is applied to the water
entry with constant vertical velocity of an infinite wedge.
An example of the convergence process for a wedge with 10 degrees
deadrise angle is shown in Fig.~\ref{conv_10}. 

It is worth noticing that for convenience, in the wedge entry problem, 
the first guess for the free surface is derived from the
dipole solution at the far field. 
Indeed, if $C_D \varphi_D$ approximates the far field behaviour, then 
the free surface elevation should behave as
\begin{equation}
\label{fs0}
\eta(0) = \frac{C_D}{3 \xi^2} \;\;,
\end{equation}
where the constant $C_D$ is derived together with the solution of the
boundary value problem, as discussed in section 3.
Note that, a few preliminary iterations are performed in order to
get a better estimate of the constant $C_D$. During these preliminary
iterations the free surface varies only due to the variation of the
coefficient $C_D$ in equation (\ref{fs0}).

Once the pseudo time stepping procedure starts, it gradually develops a
thin jet along the body surface. When the angle formed by the free
surface with the body surface drops below a threshold value, usually 10
degrees, the jet is truncated or the shallow water model is activated.
This process is shown in the left picture of Fig.~\ref{conv_10} where,
for the sake of clarity, the shallow water region is not displayed.

\begin{figure}
\centerline{\hbox{
\includegraphics[width=80mm]{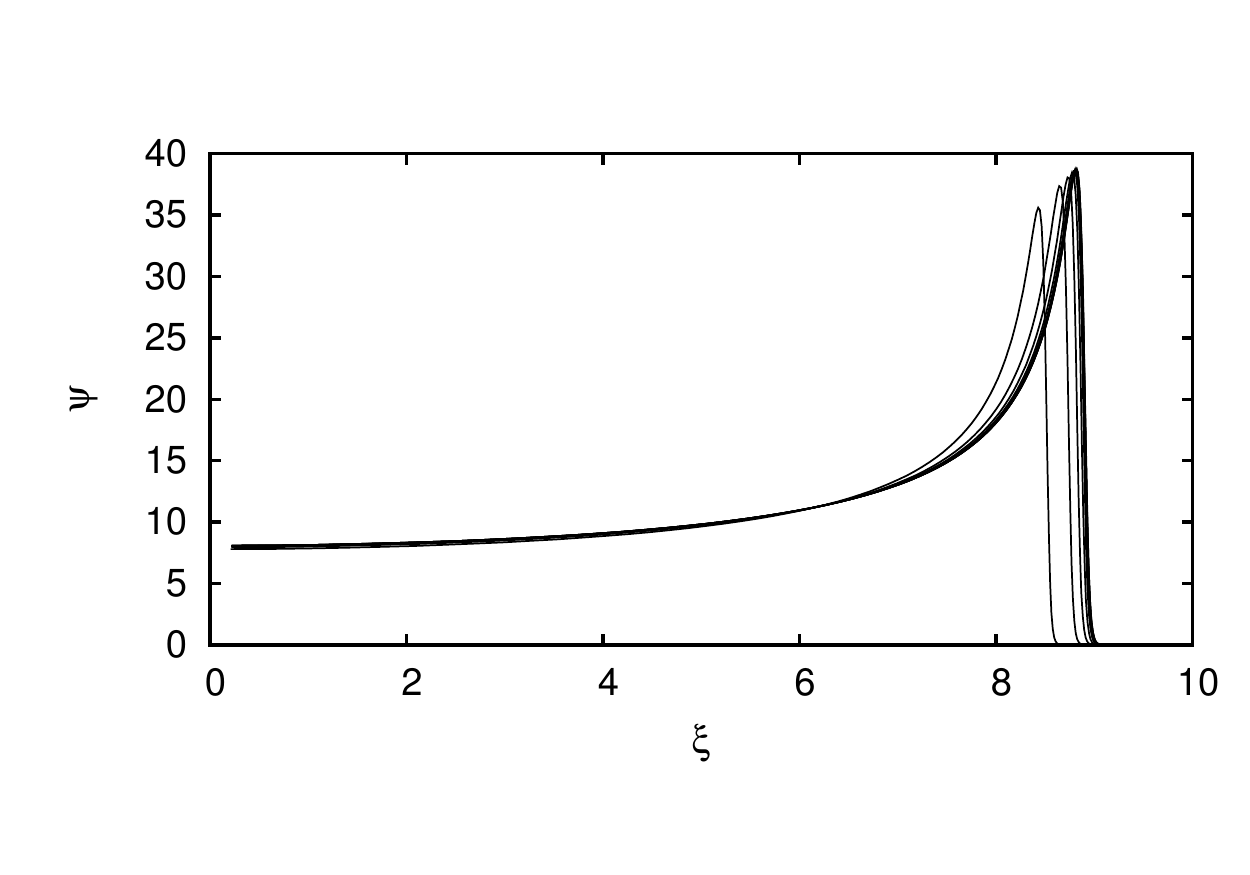}
\includegraphics[width=80mm]{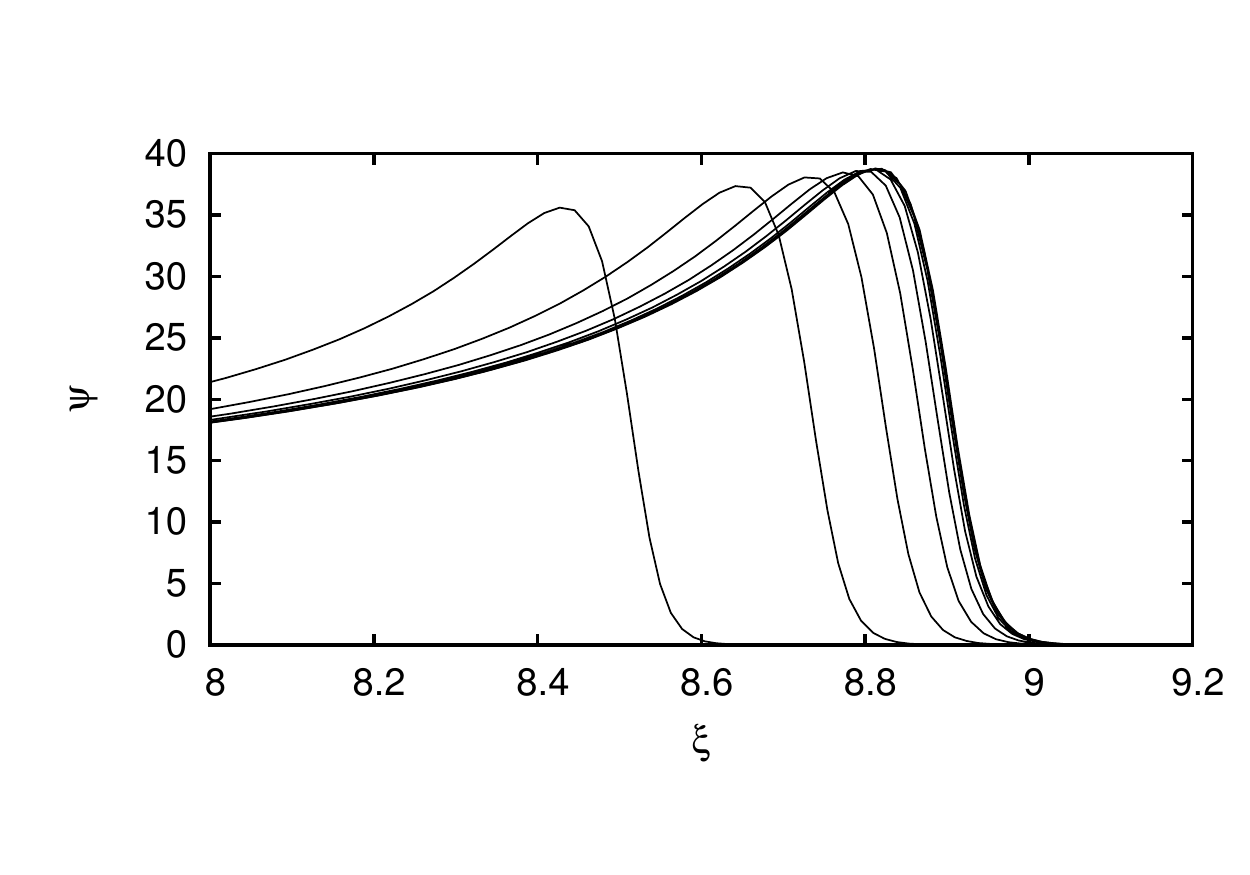}
}}
\caption{Convergence of the pressure distribution for the 10 degrees
wedge. A close up view of
the pressure about the jet root is shown, displaying a smooth
transition to zero thanks to the use of the shallow water model.\label{conv_pre}}
\end{figure}

The convergence in terms of pressure distribution is shown in
Fig.~\ref{conv_pre} along with a close up view of the jet root region. 
It can be noticed that the use of the shallow water solution within the
boundary integral representation makes the pressure very smooth 
about the transition.

\begin{figure}
\centerline{\hbox{
\includegraphics[width=80mm]{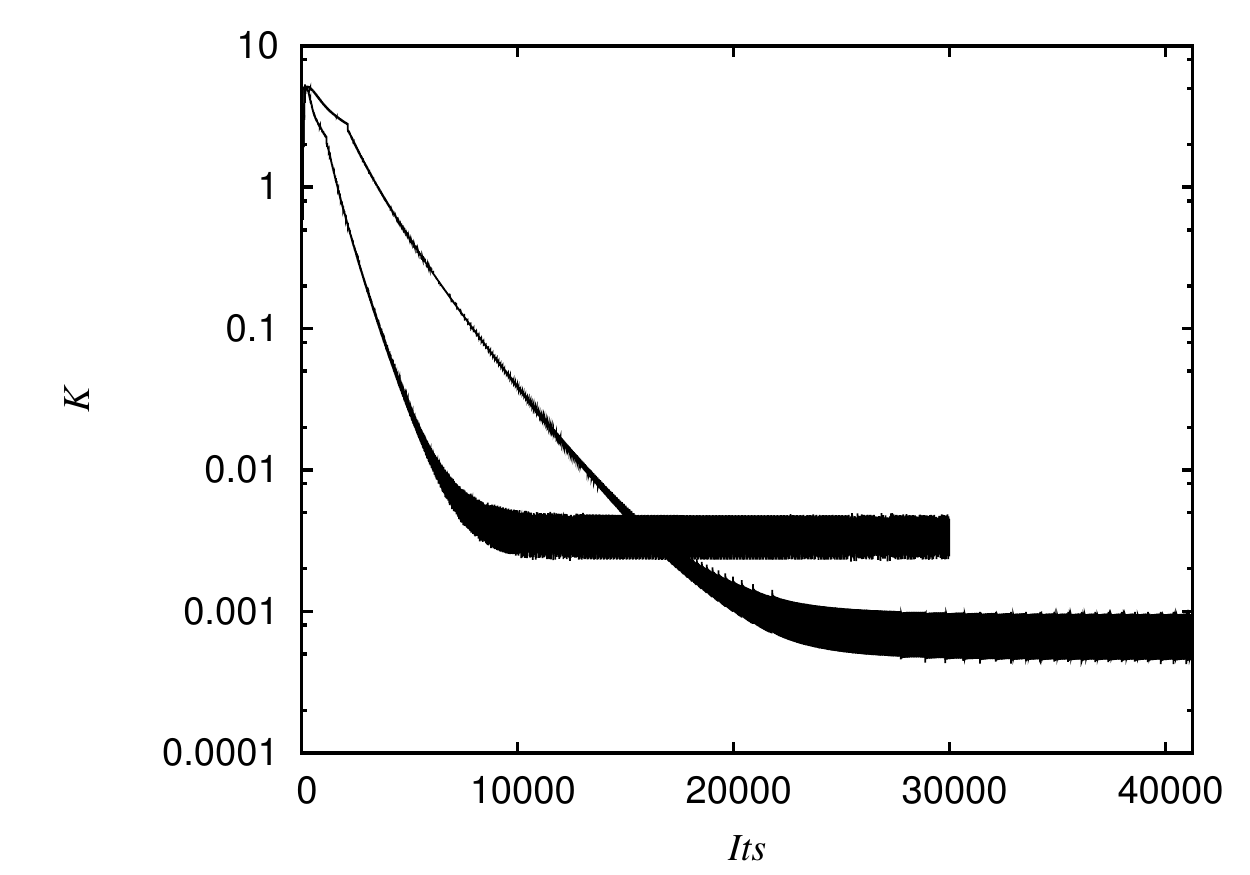}
}}
\caption{Convergence history of the kinematic boundary condition. The
curves refer to two different discretizations, the coarser having a
minimum panel size of 0.04, the finer 0.01.\label{conv_kbc}}
\end{figure}

Due to the use of the $\nabla S$ as a pseudo velocity field for the
displacement of the free surface panels, the achievement of convergence 
implies that there is no further motion of the free surface in the normal 
direction, and thus the kinematic condition (\ref{kbcssm}) is satisfied.
A more quantitative understanding of the convergence in terms of the
kinematic condition is provided by the quantity
\begin{equation}
K = \int_{S_S} S_\nu^2 \mbox{ d}s \;\;.
\end{equation}
The behaviour of $K$ versus the iteration number is displayed in
Fig.~\ref{conv_kbc}, which indicates that $K$ diminishes until reaching a
limit value that diminishes when refining the discretization.

\begin{table}
\begin{center}
\begin{tabular}{|c|c|c|c|c|} \hline
$\gamma$ & \multicolumn{2}{|c|}{$C_p=2 \psi$} & \multicolumn{2}{|c|}
{$\eta_{max}$} \\ \cline{2-5}
~ & Present & ZF & Present & ZF \\ \hline
7.5 & 140.1 & 140.59 & 0.5601  & 0.5623 \\ \hline
10 & 77.6 & 77.85 & 0.5538  & 0.5556 \\ \hline
20 & 17.7 & 17.77 & 0.5079  & 0.5087 \\ \hline
30 & 6.86 & 6.927 & 0.4217  & 0.4243  \\ \hline
40 & 3.23  & 3.266 & 0.2946  & 0.2866 \\ \hline
\end{tabular}
\caption{Comparison between the results provided by the present solver
and the corresponding data derived by the self-similar solution in
\cite{zhao93}. \label{table-sol}}
\end{center}
\end{table}

\begin{figure}
\centerline{\hbox{
\includegraphics[width=130mm]{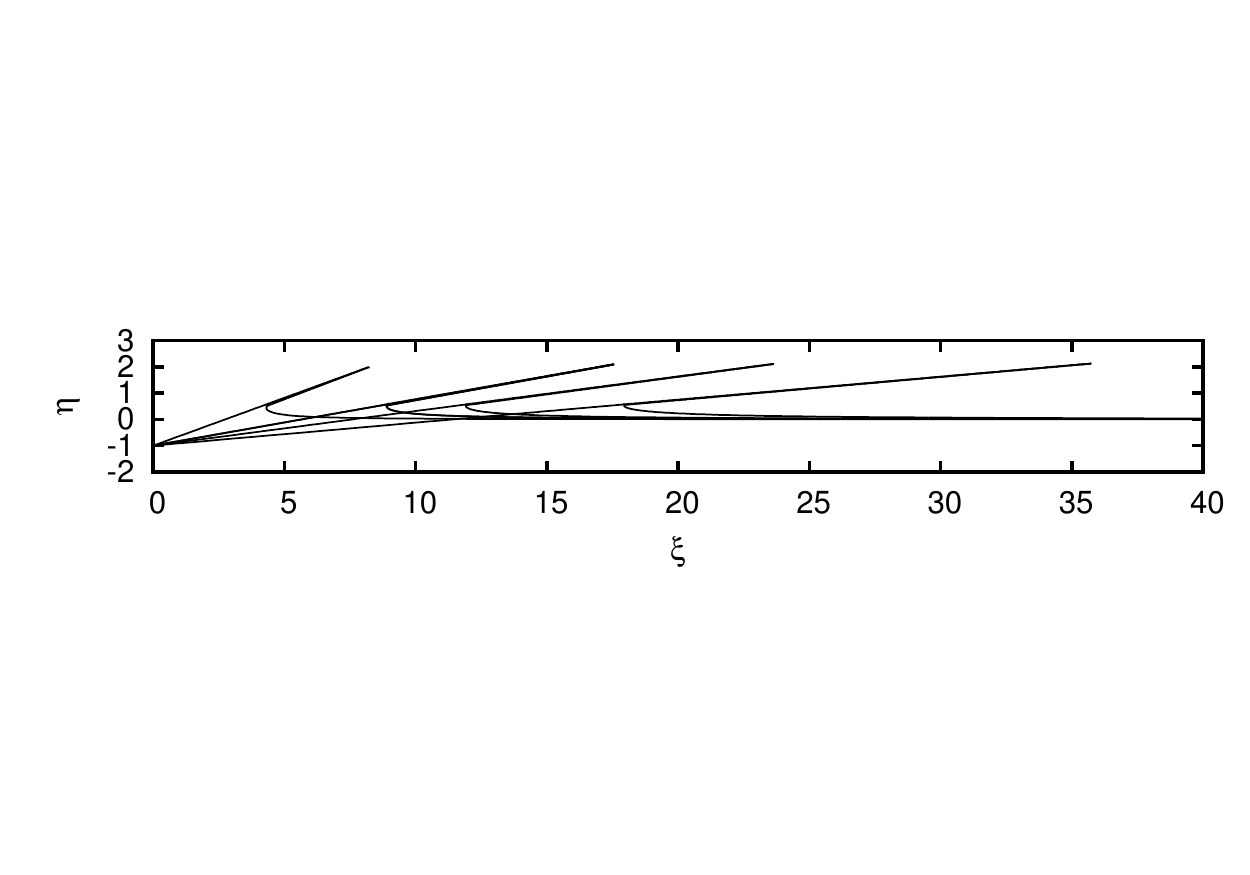}
}}
\centerline{\hbox{
\includegraphics[width=80mm]{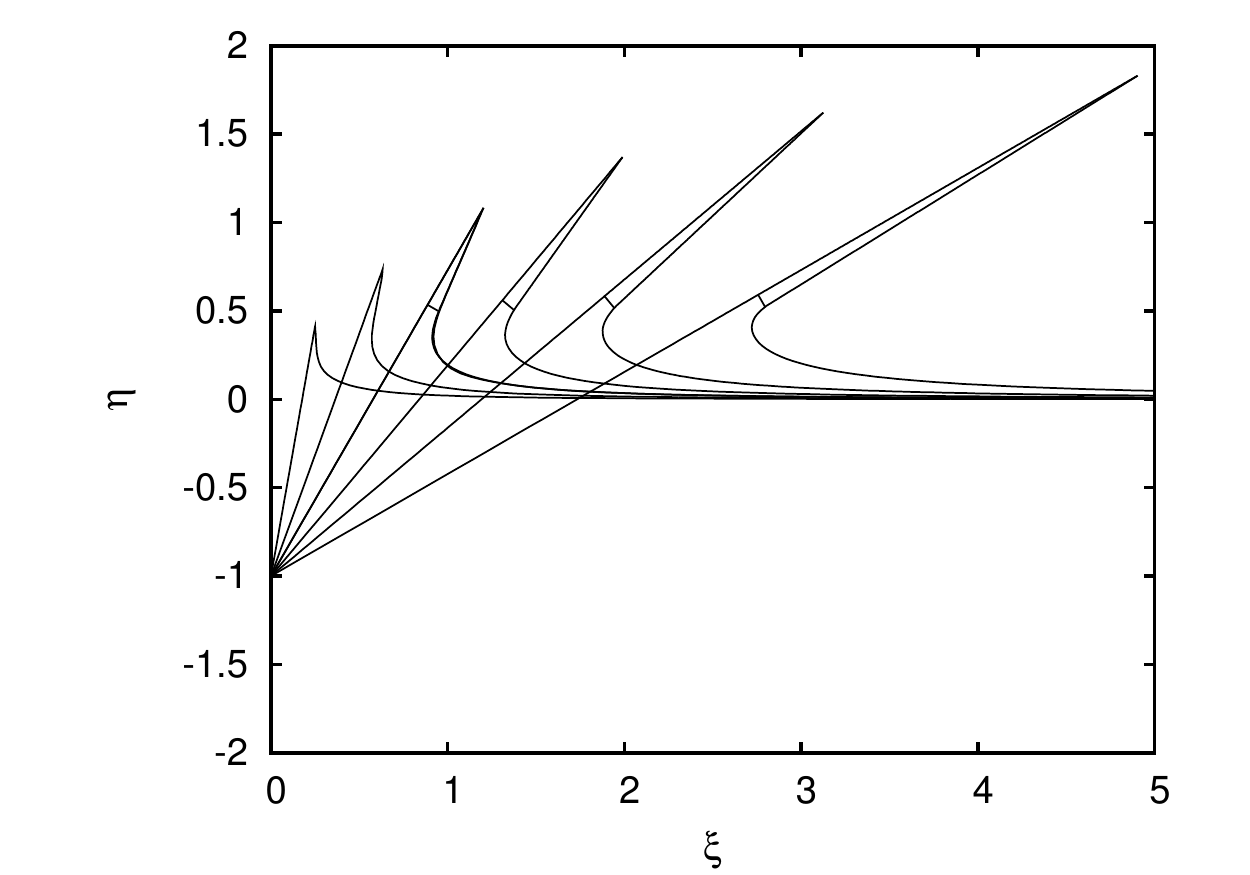}
}}
\caption{Comparison between the free surface profiles. In the upper
picture, solutions refer to $5, 7.5, 10, 20$ degrees deadrise angle. In lower
picture, solution refer to $30, 40, 50, 60, 70 $ and $80$ degrees
deadrise angle.\label{comp_cfg}}
\end{figure}

In Fig.~\ref{comp_cfg} the free surface profiles obtained
for different deadrise angles are compared.
In the figure, two solutions, largely overlapped, are drawn
for the 10 degree wedge which refers to two different discretizations,
with minimum panel size of 0.04 and 0.01 for the coarse and fine grids,
respectively.
Similarly, two solutions are drawn for the 60 degrees case, one which
makes use of the shallow water model and a second solution in which the
jet is described by the boundary integral representation. At such
deadrise angles, the angle at the tip, which is about 15 degrees, is
large enough to allow an accurate and still efficient description of the 
solution within the standard boundary integral representation.

\begin{figure}
\centerline{\hbox{
\includegraphics[width=80mm]{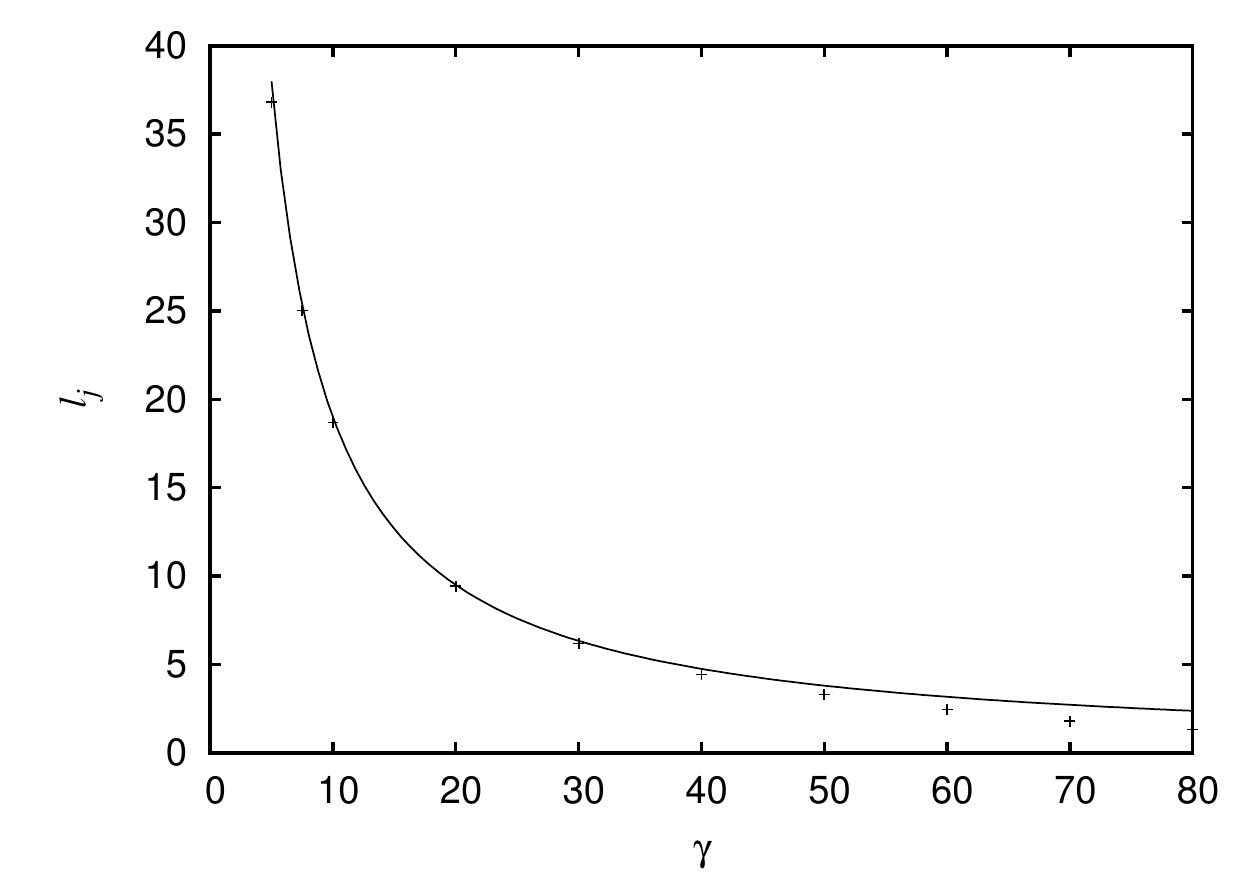}
\includegraphics[width=80mm]{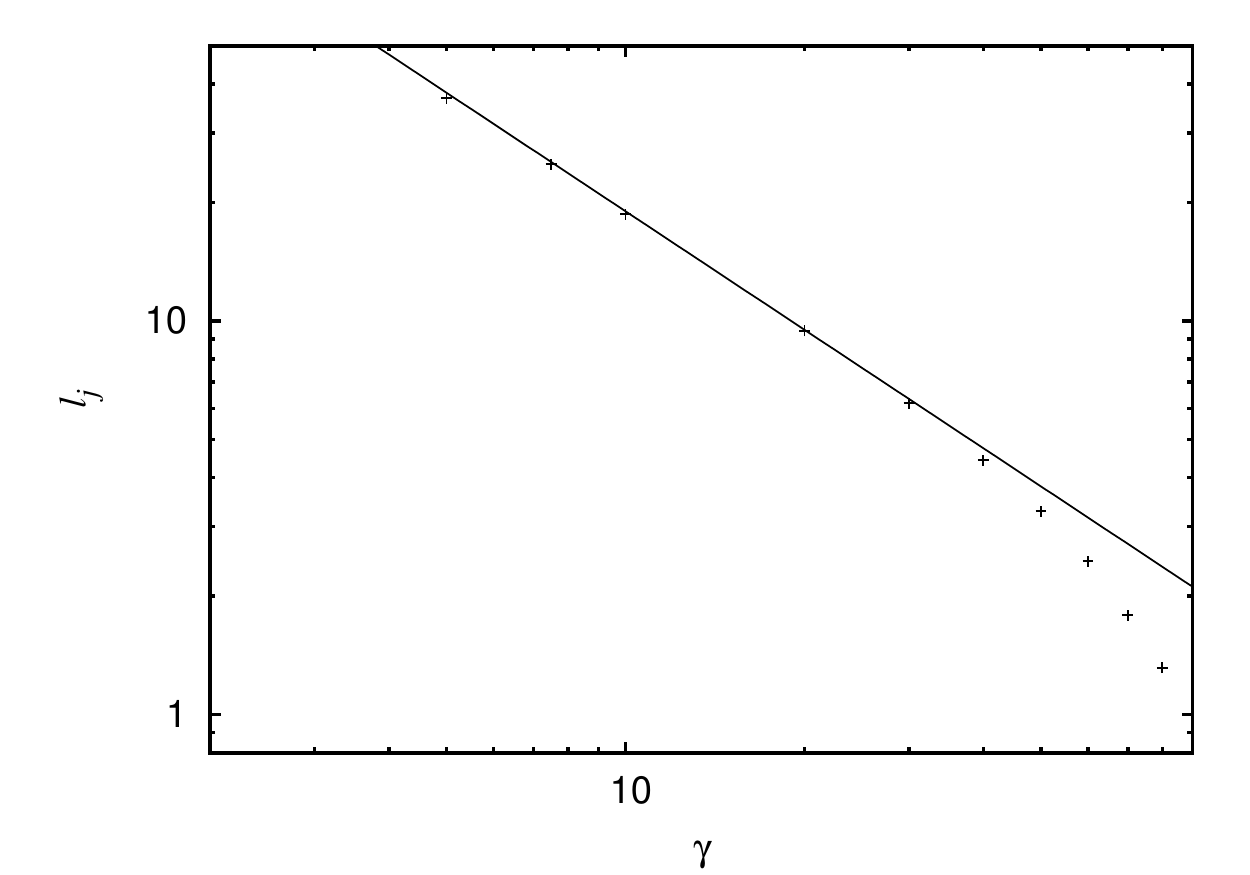}
}}
\caption{Distance of the jet tip to the wedge apex, versus the wedge
angle. According to the definition of self-similar variables
(\ref{selfvar}), $l_j V$ represents the velocity at which the
tip swept the body surface. The dash line represent the $C/\gamma$ line,
where $C=190$ if the angle is expressed in degrees. The graph is given
in logscale on the right.\label{jetlen}}
\end{figure}

It is worth noticing that, as the solution is given in terms of the
self-similar variables (\ref{selfvar}), the length of the jet in
terms of those variables
\begin{equation}
l_j = \sqrt{ \xi_j^2+\eta_j^2} \;\;,
\end{equation}
is also an index of the propagation velocity of the tip. For the present
problem, according to equations (\ref{selfvar}) the scaling from the 
self-similar to the physical variables is simply linear and thus the
$l_j$ is just the tip speed.
The jet length versus the deadrise angle, which is drawn in Fig.~\ref{jetlen},
approaches a $1/\gamma$ trend for $\gamma \le 30$ degrees.

\begin{figure}
\centerline{\hbox{
\includegraphics[width=85mm]{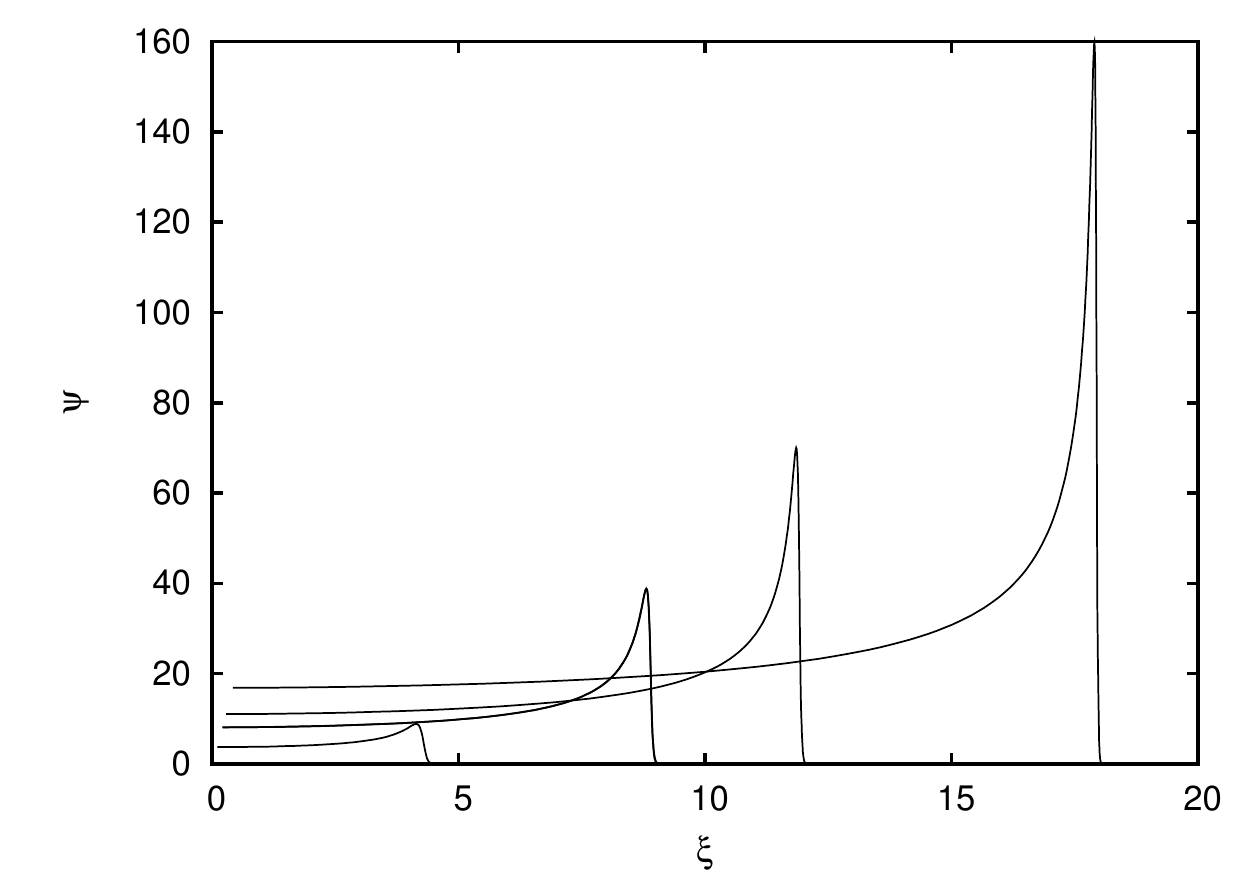}
\includegraphics[width=85mm]{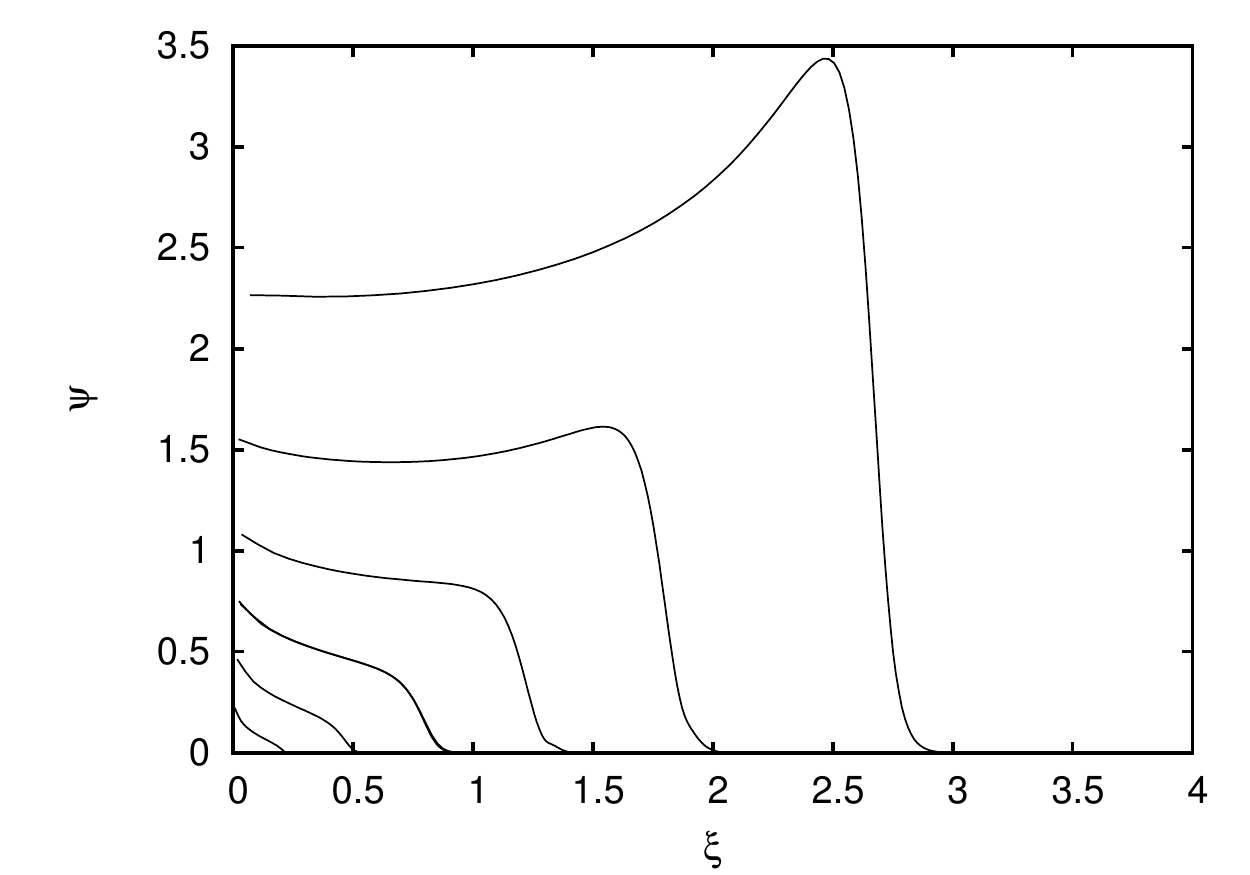}
}}
\caption{Comparison between the pressure distributions. Solutions for
$5, 7.5, 10, 20$ degrees deadrise angle are shown on the left, whereas
solutions for $30, 40, 50, 60, 70 $ and $80$ degrees deadrise angle are
given on the right.\label{comp_pre}}
\end{figure}

A comparison of the pressure distribution obtained for different
deadrise angles is provided in Fig.~\ref{comp_pre}.
The results show that, up to 40 degrees deadrise angle,
the pressure distribution is characterised by a pressure peak occurring
about the root of the jet whereas, at larger deadrise angles, the maximum
pressure occur at the wedge apex.
As for the free surface shape, also for the pressure distribution two
solutions are drawn for the cases at 10 and 60 deadrise angles, which are
essentially overlapped.

For validation of the results, some relevant quantities are extracted and 
compared with corresponding data computed by the \cite{dobro69} model in 
\cite{zhao93}.
The comparison is established in terms of the maximum
pressure coefficient, which is $2 \psi_{max}$, and of the vertical
coordinate of the point along the body where the pressure gets the peak,
$\eta_{max} = \xi_{max} \tan \gamma - 1$.
The comparison, shown in Table~\ref{table-sol},
displays a rather good agreement for all the deadrise angles.

\begin{figure}
\centerline{\hbox{
\includegraphics[width=80mm]{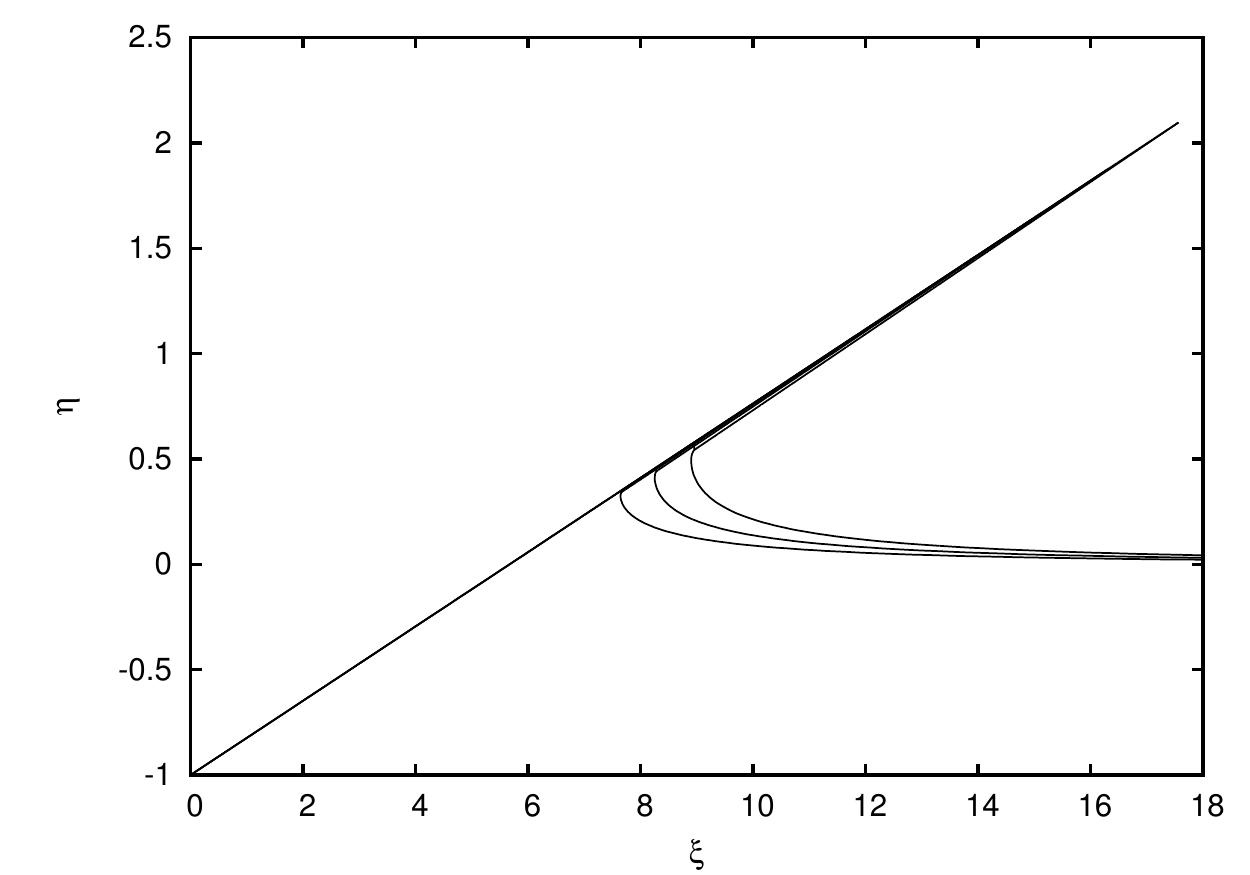}
\includegraphics[width=80mm]{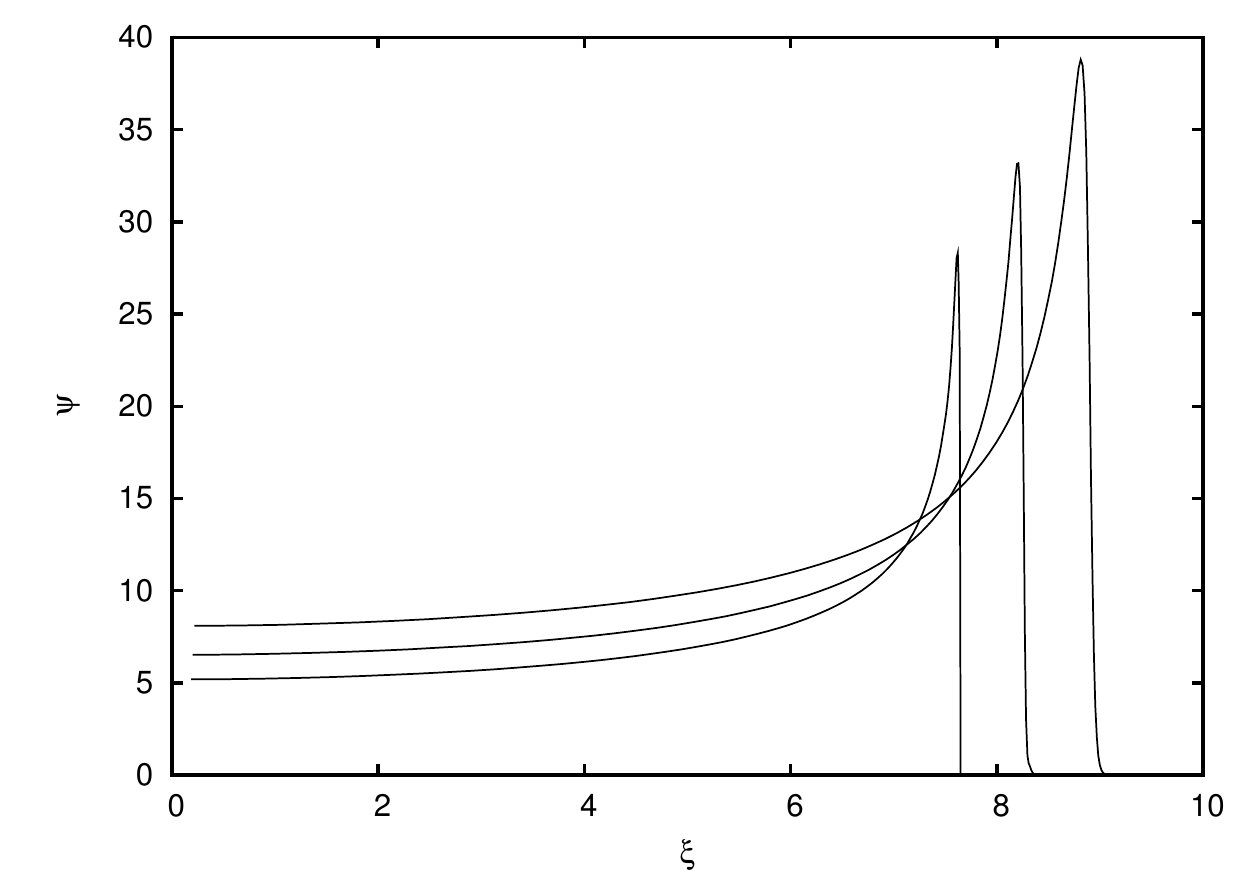}
}}
\caption{Effect of the porosity on the free surface shape and pressure
distribution. 
Solutions are drawn for deadrise angle $\gamma=10$ degree, with porosity
coefficients $\alpha_0=0, 0.02 $ and $0.05$. 
\label{cfgpor10}}
\end{figure}

When the body surface is either porous or perforated, a flow through the
solid boundary occurs which grows with the local pressure.
As shown in Fig.~\ref{comp_pre}, for deadrise angles smaller than 40
degrees the pressure takes the maximum at the root of the jet, and this
causes a shrinking, and a subsequent shortening, of the jet.
The changes in the free surface shape caused by the porosity of the
surface on a 10 degrees wedge are shown in Fig.~\ref{cfgpor10} where the
shrinking of the jet is clearly highlighted.
From the corresponding pressure distributions, which are given in the
left picture, it can be seen that even low porosity levels provide an
important reduction in the pressure peak, and the peak itself is shifted 
down towards the wedge apex, thus leading to a shortening of the region
 of the body exposed to the pressure.
Beside the reduction of the peak value, the pressure displays a significant
reduction also in the remaining part of the body, and all those effects
combined together implies an significant reduction of the total load 
acting on the body.

\begin{figure}
\centerline{\hbox{
\includegraphics[width=80mm]{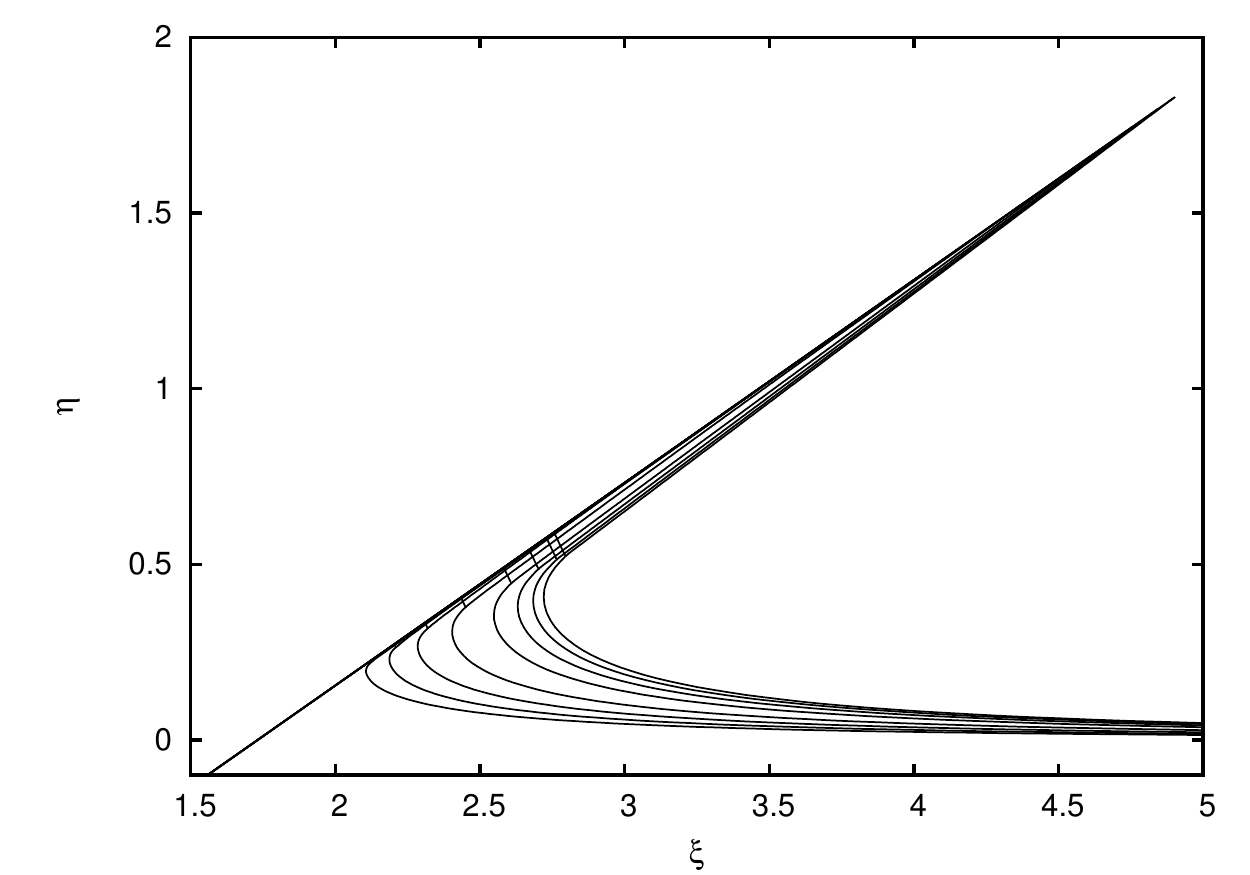}
\includegraphics[width=80mm]{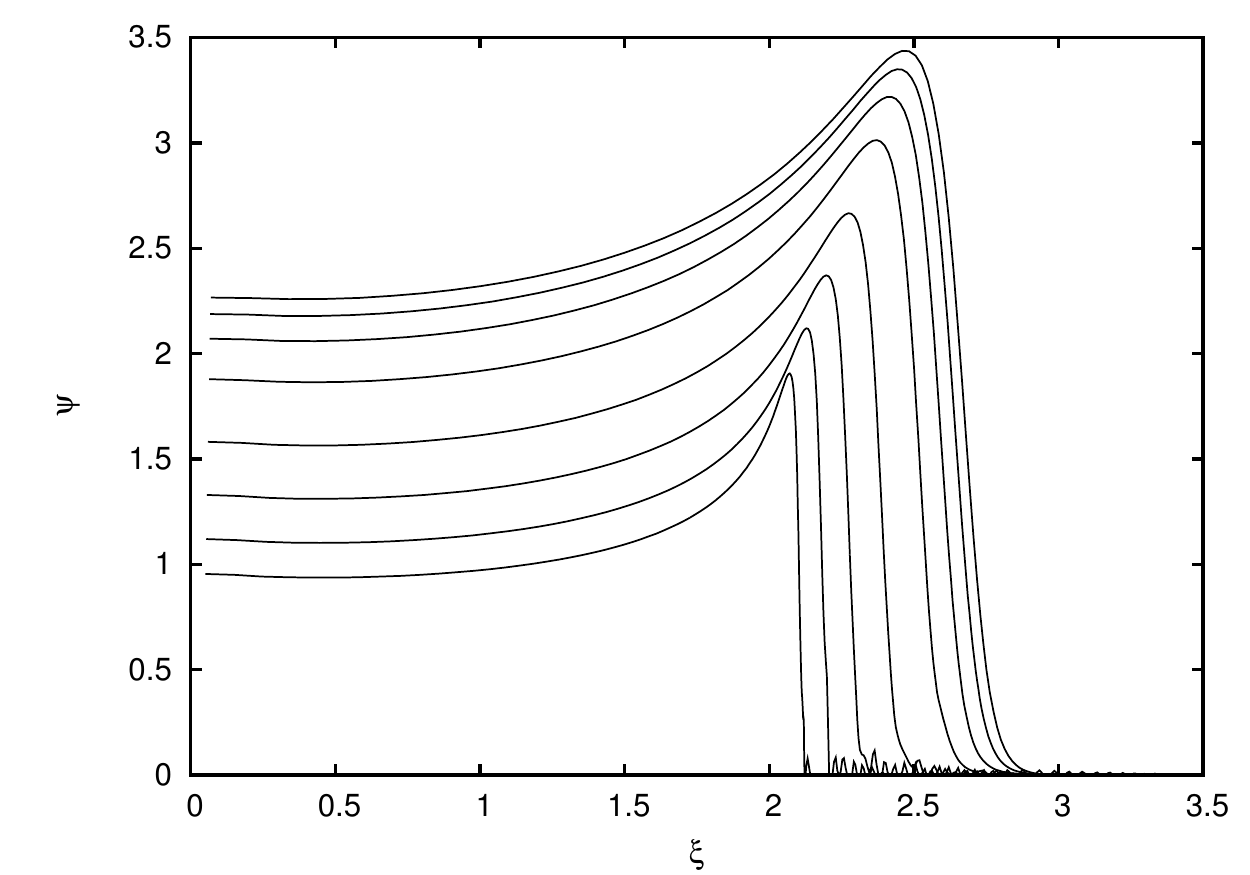}
}}
\caption{Free surface shapes and pressure distributions for a wedge 
$\gamma=30$ degrees deadrise angle in case of perforated
surfaces. The perforated coefficients are $0, 0.02, 0.05, 0.10, 0.20, 
0.30, 0.40, 0.50$.\label{cfgpre30}}
\end{figure}

Similar results are shown also for a 30 degrees wedge, in
Fig.~\ref{cfgpre30}, where solutions for perforated surfaces are 
compared. Quantitatively, the differences in terms of $C_p$, $\eta_{max}$ 
and $l_j$ when varying the porous or the perforation coefficient are 
provided in Table~\ref{table_por}.

\begin{table}
\begin{center}
\begin{tabular}{|c|c|c|c|c|c|c|} \hline
$\alpha_0, \chi$ & \multicolumn{3}{|c|}{Porous} & \multicolumn{3}{|c|}
{Perforated} \\ \cline{2-7}
~ & $C_{p_{max}}$ & $\eta_{max}$  & $l_j$  & $C_{p_{max}}$ & 
$\eta_{max}$ & $l_j$  \\ \hline
0.0 & 6.86 & 0.4217 & 5.6612  & 6.86 & 0.4217 & 5.6612 \\ \hline
0.02 & 6.59 & 0.4052 & 5.5527  & 6.69 & 0.4098 & 5.5930 \\ \hline
0.05 & 6.20 & 0.3807 & 5.4086 & 6.43 & 0.3988 & 5.4910 \\ \hline
0.10 & 5.68 & 0.3451 & 5.2084 & 6.02 & 0.3697 & 5.3350 \\ \hline
0.20 & 4.88 & 0.2887 & 4.9145 & 5.33 & 0.3127 & 5.0610 \\ \hline
0.30 & 4.28 & 0.2442 & 4.7100 & 4.74 & 0.2686 & 4.8339 \\ \hline
0.40 & 3.83 & 0.2150 & 4.5610 & 4.24 & 0.2285 & 4.6462 \\ \hline
0.50 & 3.45 & 0.1856 & 4.4460 & 3.81 & 0.1956 & 4.4912 \\ \hline
\end{tabular}
\caption{Effect of the porosity coefficient and of the perforation ratio
on some relevant parameters for a 30 degrees wedge. \label{table_por}}
\end{center}
\end{table}

\subsection{Sliding block}

\begin{figure}[tbp]
\centerline{\hbox{
\includegraphics[width=80mm]{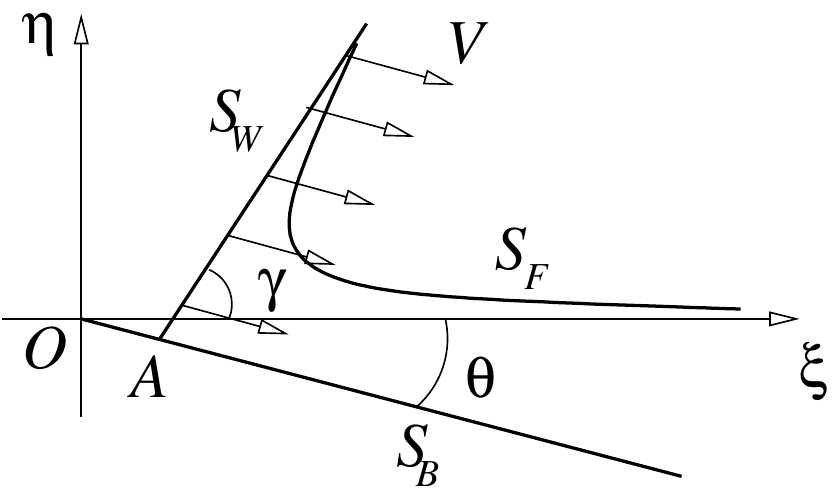}
}}
\caption{Block sliding along an inclined sloping bed.\label{scksli}}
\end{figure}

As a second application, the method is adopted to study the flow
generated when a solid block slides along a sloping bed
(Fig.~\ref{scksli}).
This flow configuration resembles that generated at coastal sites when
massive land slides along the sea bed giving rise to tsunamis. The study
can help in understanding which are the conditions in terms of the angle
of the front and bed slope which result in the larger velocities.
The possibility of accounting for the permeability of the mass
is exploited as well.

This problem and all the physical implications were already addressed and 
discussed in \cite{iafrati07b}. As explained in the introduction, the 
application is here shortly reviewed, focusing the attention on the changes 
to be operated to the boundary value problem formulated above. Some other 
aspects of the solutions are highlighted as well.

By assuming a constant entry velocity, which is acceptable in an early
stage, the flow is self-similar and can be described by the same approach
presented above.
The only difference concerns the boundary conditions on the bed and on
the front which are
\begin{equation}
\varphi_\nu = 0 \qquad \mbox{ on } \; \eta = -\xi \tan \theta
\end{equation}
and
\begin{equation}
\varphi_\nu = \sin(\gamma+\theta) -f(\psi) \qquad \mbox{ on }  \; \eta =
\xi \tan \gamma - \frac{\sin(\gamma+\theta)}{\cos \gamma}
\end{equation}
where, as aforesaid, $f(\psi)$ accounts for the permeability of the block 
and $\theta$ is the inclination of the sea bed (Fig.~\ref{scksli}). 
In this case the solid boundary is represented by both the bed and the
block front, but only the latter can be permeable.

\begin{figure}[tbp]
\centerline{\hbox{
\includegraphics[width=80mm]{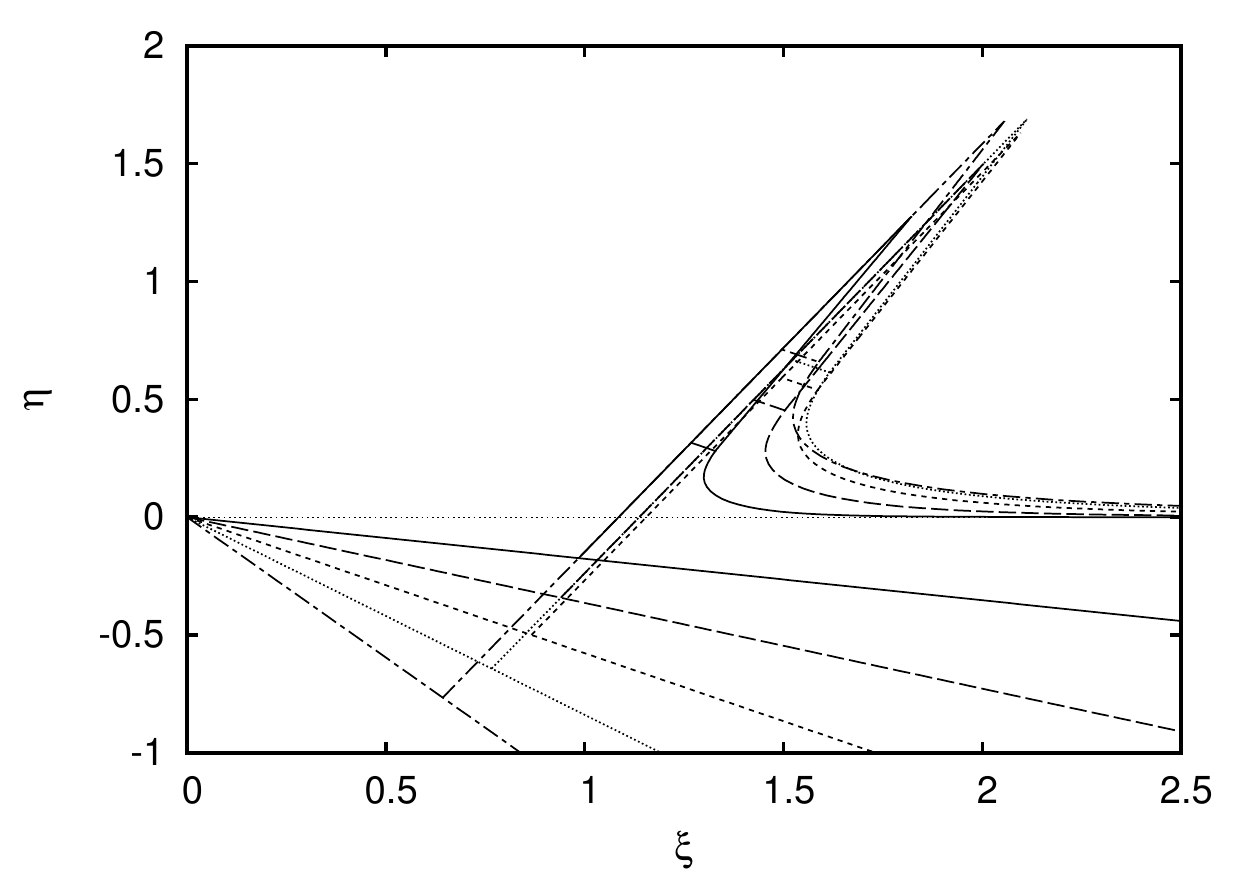} }}
\centerline{\hbox{
\includegraphics[width=80mm]{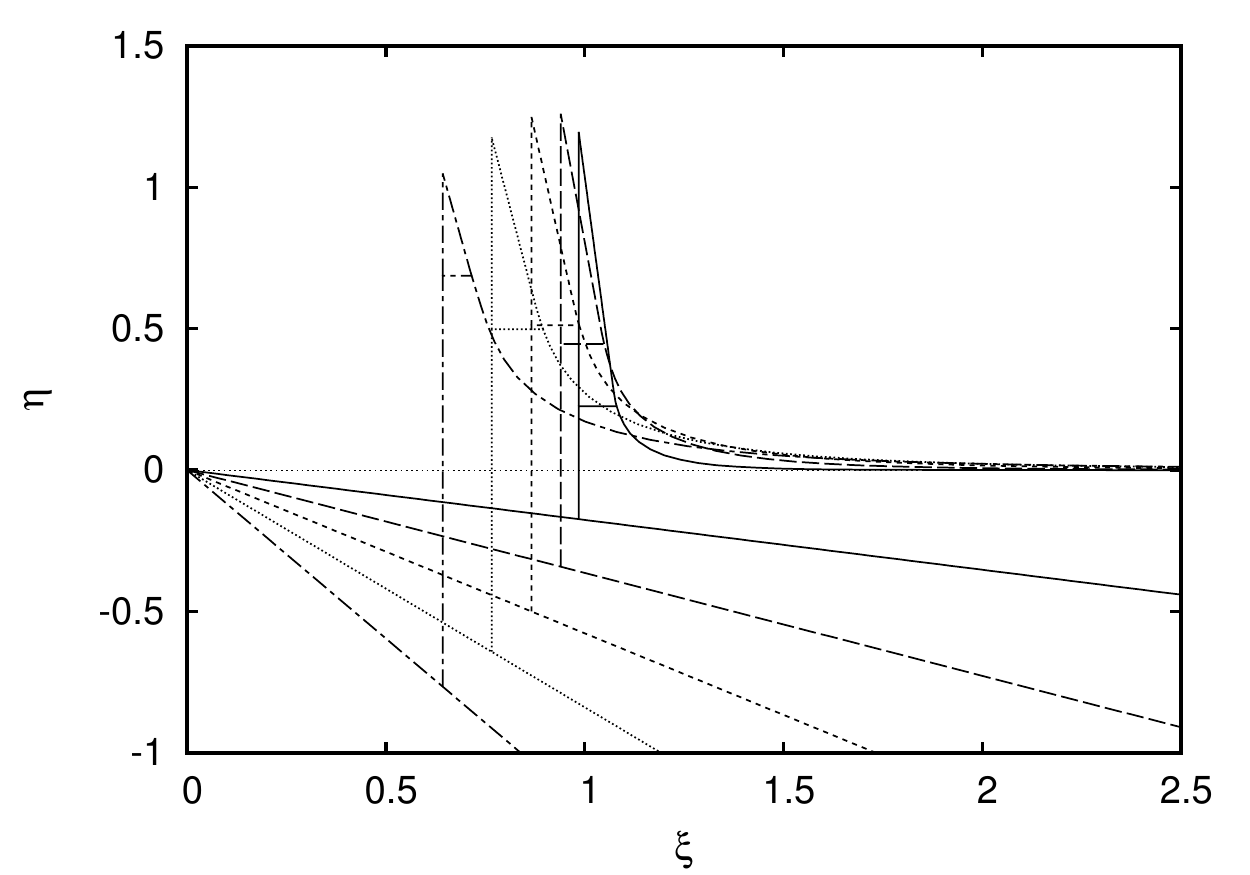} }}
\centerline{\hbox{
\includegraphics[width=80mm]{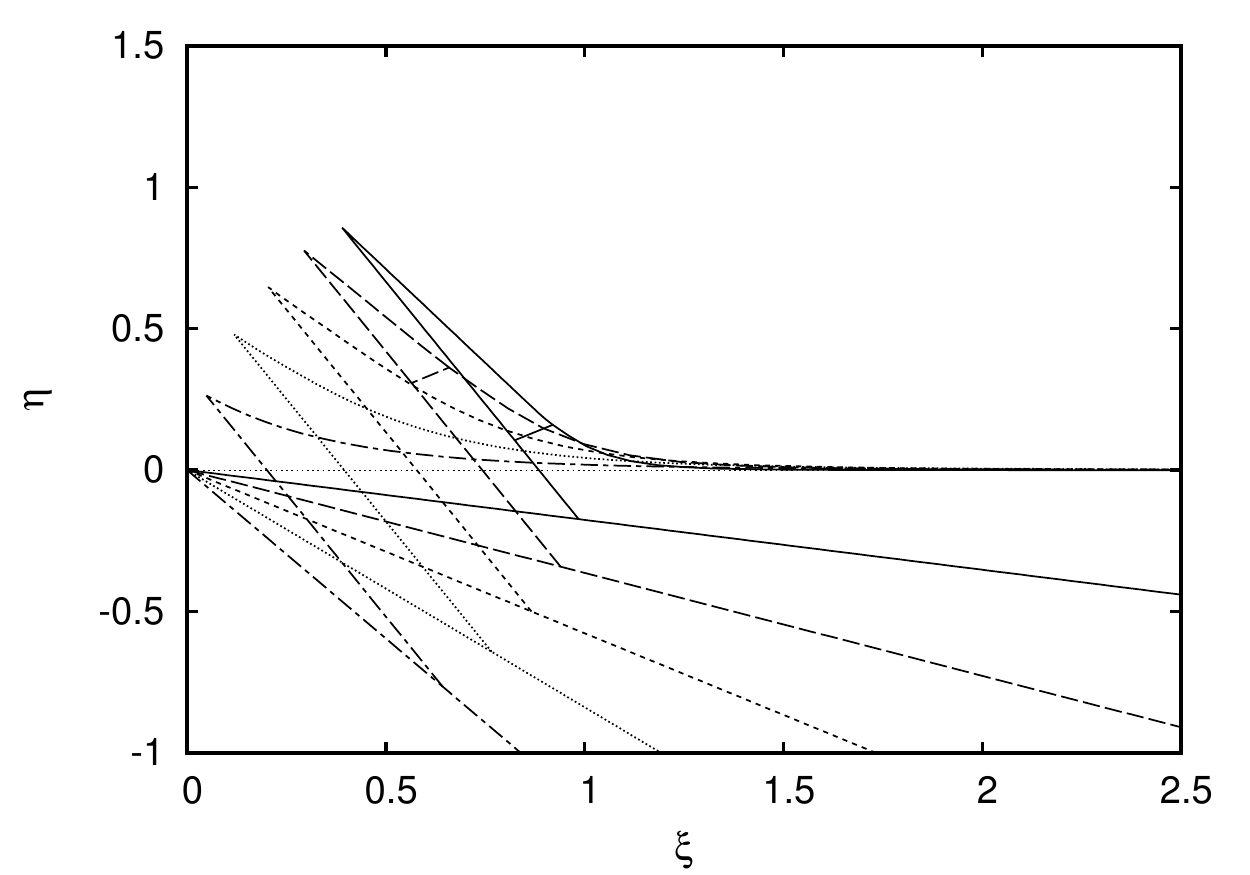}
}}
\caption{Free surface profiles for a block, 60 degrees ({\em top}), 90
degrees ({\em middle}), 120 degrees ({\em bottom}).
Solutions are drawn for bed slopes 10, 20, 30, 40 and 50 degrees. The line
about the jet root indicates the position where the shallow water model
is started.\label{cfg_sli}}
\end{figure}

The free surface profiles generated by a block sliding over sea beds with
different slopes and different inclinations of the block are shown in 
Fig.~\ref{cfg_sli}. 
When the shallow water model is exploited, the free surface portions 
belonging to the bulk of the fluid and to the shallow water region are 
displayed, along with the boundary between the two domains.
For the case with $\gamma=60$ degrees, the results indicate that
the jet length grows, i.e. the tip moves faster, when the beach slope
increases from 10 to 40 and decays for larger slopes. Results are similar
for the case $\gamma = 90$ degrees, although the maximum is achieved for a bed
slope of $30$ degrees. For an inclination of the front of $120$ degrees
the results show that the jet length decays monotonically as the beach
slope increases.
Quantitatively, the results are summarized in Table~\ref{table_sli}.

\begin{table}
\begin{center}
\begin{tabular}{|c|c|c|c|c|c|c|c|c|c|} \hline
$\theta$ & \multicolumn{3}{|c|}{$\gamma=60$} &
\multicolumn{3}{|c|}{$\gamma=90$} & \multicolumn{3}{|c|}{$\gamma=120$}
\\ \cline{2-10}
~ & $\xi_T$ & $\eta_T$ & $l_J$ & $\xi_T$ & $\eta_T$ & $l_J$ & $\xi_T$ &
$\eta_T$ & $l_J$ \\ \hline
10  & 1.822  & 1.277  & 1.474 & 0.9848 & 1.196   & 1.196 & 0.3894     &
0.8576 &  1.232 \\ \hline
20  & 2.003  & 1.500  & 1.732 & 0.9397 & 1.261   & 1.261 & 0.2940     &
0.7764 &  1.163 \\ \hline
30  & 2.097  & 1.631  & 1.884 & 0.8660 & 1.249   & 1.249 & 0.2039     &
0.6468 &  0.8670 \\ \hline
40  & 2.112  & 1.689  & 1.951 & 0.7660 & 1.176   & 1.176  & 0.1182     &
0.4794 &  0.6211 \\ \hline
50  & 2.056  & 1.682  & 1.942 & 0.6428 & 1.050   & 1.050 & 4.8543E-02 &
0.2632 &  0.3309 \\ \hline
60  & 1.931  & 1.613  & 1.863 & 0.5000 & 0.8749  & 0.8749 & ~ & ~&~ \\
\hline
\end{tabular}
\caption{Coordinates of the jet tip $(\xi_T,\eta_T)$ and length of the 
jet $l_J$ for a block with $\gamma = 60, 90 $ and $120$ degrees sliding 
along a seabed with different slopes. \label{table_sli}}
\end{center}
\end{table}

\begin{figure}[p]
\centerline{\hbox{
\includegraphics[width=80mm]{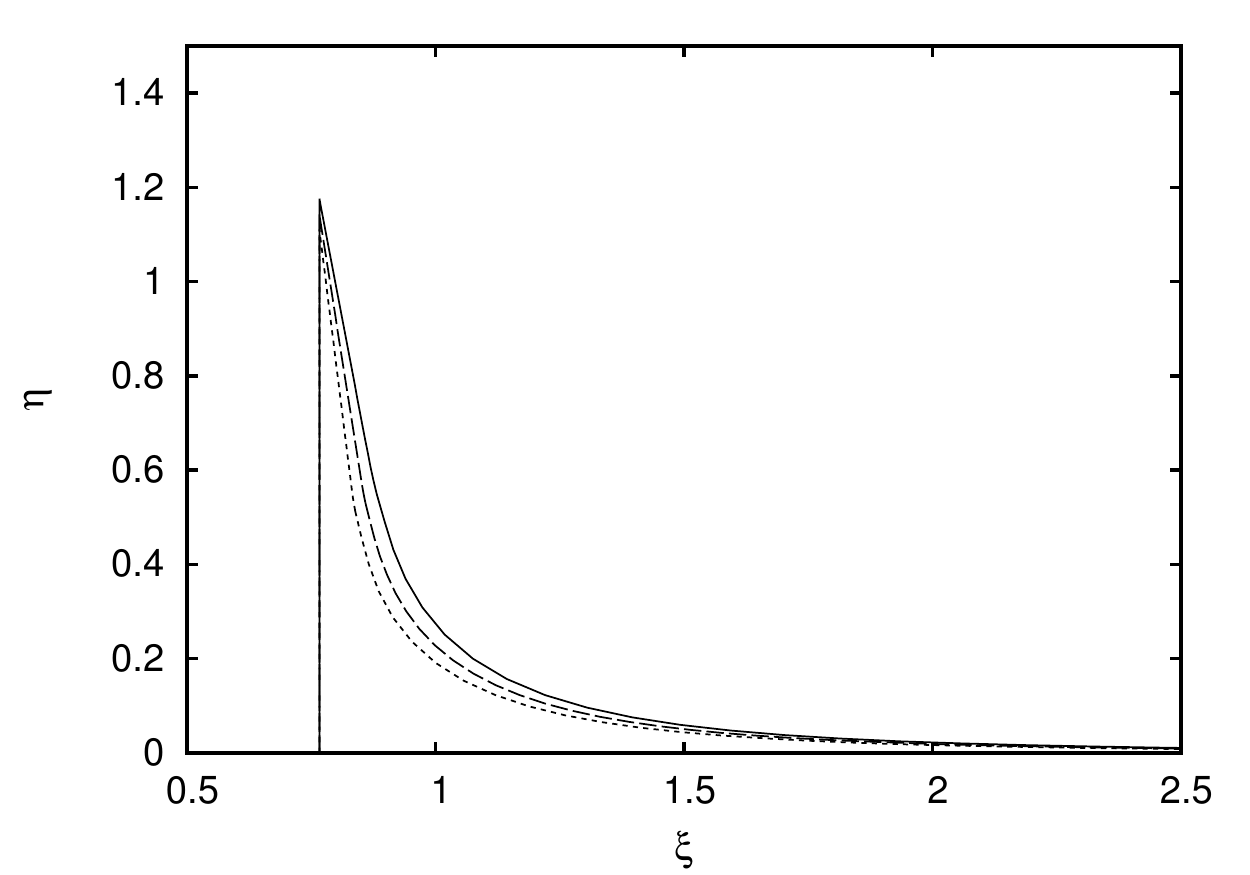}
}}
\caption{Effect of the permeability of the front on the free surface
elevation. Solutions refer to a block front 90 degrees inclination, with
$\chi = 0$ ({\em solid}), $\chi = 0.1$ ({\em dash}) and $\chi = 0.2$
 ({\em dot}).\label{cfg_sli_por}}
\end{figure}

In the case of a permeable front, the flow across the solid boundary
makes the jet thinner and shorter, i.e. the tip speed is lower.
This can be seen from Fig.~\ref{cfg_sli_por} where the free surface profiles
obtained for a perforated block with $\gamma = 90$ degrees and $\theta = 40$
are shown for $\chi = 0, 0.1$ and $0.2$.

\subsection{Floating plate impact}

The computational procedure can be also applied in contexts in which the
problem is not strictly self-similar but is can be approximated as
self-similar under additional assumptions.
This is for example the case of the sudden start of a wedge
originally floating on the free surface with the apex submerged 
\citep{iafrati05} or the sudden start of a floating wedge in a weakly 
compressible liquid \citep{korobkin06}. 
Both problems are approximately self-similar in the small time limit.
It can be shown that the problems can be represented by the same boundary 
value problem, aside from some differences in the coefficients appearing 
in the dynamic boundary condition (\ref{dbcss}) or (\ref{dbcssf}).

As a particular example, the water entry of a floating plate is presented
here below.
Formally, the plate entry problem is not self-similar as the
breadth of the plate introduces a length scale. 
By formulating the solution of the problem in the form of a small time
expasion, it can be shown that the first order solution is singular at the 
edge of the plate. In order to resolve the singularity, an inner solution 
has to be formulated under set of stretched coordinates. 
Hence, the inner solution has to be matched to the outer one at the far field. 

It is worth remarking that a detailed derivation of the outer and inner 
problems, as well as the matching condition, can be found in \cite{iafrati04}
whereas practical applications are discussed in \cite{iafrati08} and
\cite{iafrati11}. The problem is here discussed in order to highligth
the different form of the dynamic boundary condition compared to the
previous cases. This requires a different procedure as the dynamic boundary 
condition cannot be integrated analytically and, moreover, an additional 
unknown appears
which has to be derived as a part of the solution. As shown in the
following, the additional unknown governs the shape of the free surface.

Within a small time assumption, the problem in a close neighbourood of the 
edge is formulated in terms of the following variables:
\begin{equation}
\label{selfvarp}
\xi = \frac{x-1}{B t^{2/3}} \qquad
\eta = \frac{y}{B t^{2/3}} \qquad
\varphi = \frac{\phi}{\sqrt{2B} t^{1/3}} \;\;, B=(9/2)^{1/3} \;\;.
\end{equation}
In terms of the new variables, the problem is such that the plate is 
fixed and the flow is arriving from the far field. As $t \to 0$, in
terms of the stretched coordinates (\ref{selfvarp}), the plate occupies the
negative $\xi$-axis, with the plate edge located at the origin of the
coordinate system.
With respect to the pure self-similar problem, some differences occur in 
terms of the boundary conditions, which are
\begin{eqnarray}
\label{bbcfp}
\varphi_\nu & = & 0\qquad \qquad \qquad \xi < 0, \eta=0 \\
\label{dbcssfp}
S_\tau &=& \pm \sqrt{\frac{1}{2}\rho^2-S}  \qquad\qquad h(\xi,\eta) = 0 \\
\label{kutta}
h(0,0)&=&0 \;, \; h_\xi(0,0) = 0 \qquad \qquad (\xi=0, \eta=0) \\
\label{sffp}
\varphi & \to & \sqrt{\rho} \sin(\delta/2) \qquad \qquad \rho \to \infty
\end{eqnarray}
where equations (\ref{kutta}) states that the free surface is
always attached at the plate edge, and leaves the plate tangentially
\citep{iafrati04}.

Although the solution procedure is quite similar to that presented
above, there are some important differences which deserve a deeper
discussion.
Due to the different coefficients in the dynamic boundary condition, it 
cannot be analytically integrated along the free surface. 
Equation (\ref{dbcssfp}) is integrated numerically along the free surface 
starting from the far field where the matching with the asymptotic 
behaviour of the solution is enforced.
The numerical integration needs care, in particular at large distances
from the origin, in order to avoid the effects of round off errors.
Aside from that, the problem is even more complicated as the sign
in the dynamic boundary condition (\ref{dbcssfp}) changes along the 
free surface. 
This can be easily understood by
examining the behaviour of the solution nearby the plate edge, $\rho \to
0$, and in the far field as $\rho \to \infty$. Close to the edge, the flow 
exits from the area beneath the plate toward the free surface.
If $\tau$ denotes the arclength along the free surface, oriented toward the 
far field, we have that $\varphi_\tau > 0$, and thus, as $\rho=0$ at the 
edge, $S_\tau>0$.
At the far field, the free surface approach the undisturbed water level
and thus, from the definition (\ref{sdef}) it follows that
$S_\tau \to - \rho$.

The position of the inversion point, i.e. the point where the sign in
the dynamic condition changes from negative to positive, is unknown and
has to be determined as a part of the solution. The additional
constraint given by equations (\ref{kutta}) is used to that purpose.
In discrete form, at each time step, three distributions of the velocity 
potential are initialized on the free surface by locating the inversion 
point at the same panel vertex used at the previous step, and at the 
vertices of the preceeding and successive panels.
Three boundary value problems are solved by using the three distributions
of the potential and the three values of the normal derivatives of
the velocity potential at the first panel attached at the plate  are
compared. The inversion point is located at that position among the
three, which yields the smallest value of the normal velocity at the
plate edge \citep{iafrati04}.

Physically, the inversion point represents a point of discontinuity in
the tangential velocity along the free surface. Indeed, such a point is
the tip of the thin splash developing at short distance from the edge. 
The discontinuity in the tangential velocity is associated to the 
discontinuity in the tangent to the free surface at the tip.
Due to the lack of a known surface to be used as a base for the shallow
water model, in this case the flow within the thin splash is described 
by the boundary integral representation, aside from the very thin part
which is cut off. The discretization is continuously refined in order to 
ensure an adequate resolution throughout the spray, whereas the normal
velocity at the trucation is assigned to be equal to the projection of
the velocities at the two sides of the free surface along the normal to 
the truncation panel.

\begin{figure}
\centerline{\hbox{
\includegraphics[width=80mm]{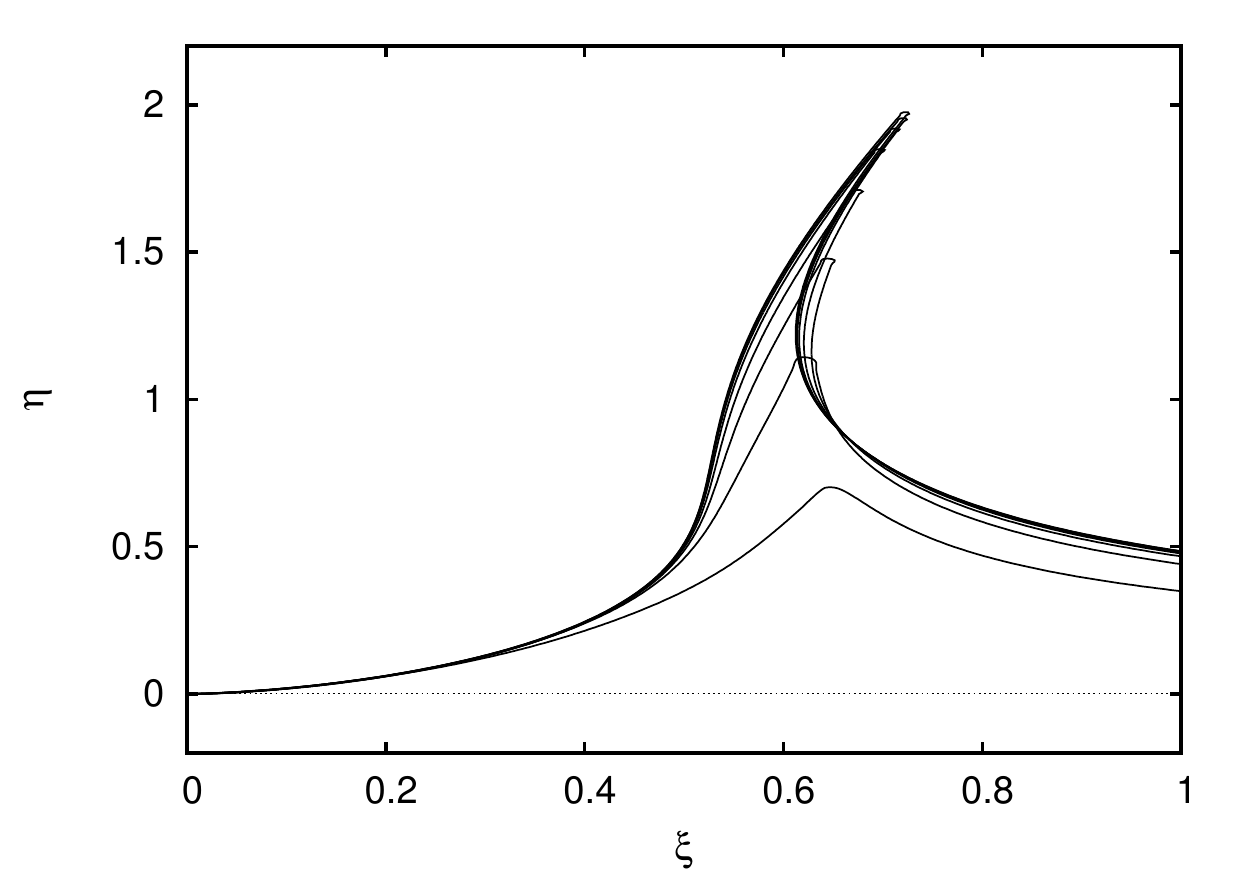}
\includegraphics[width=80mm]{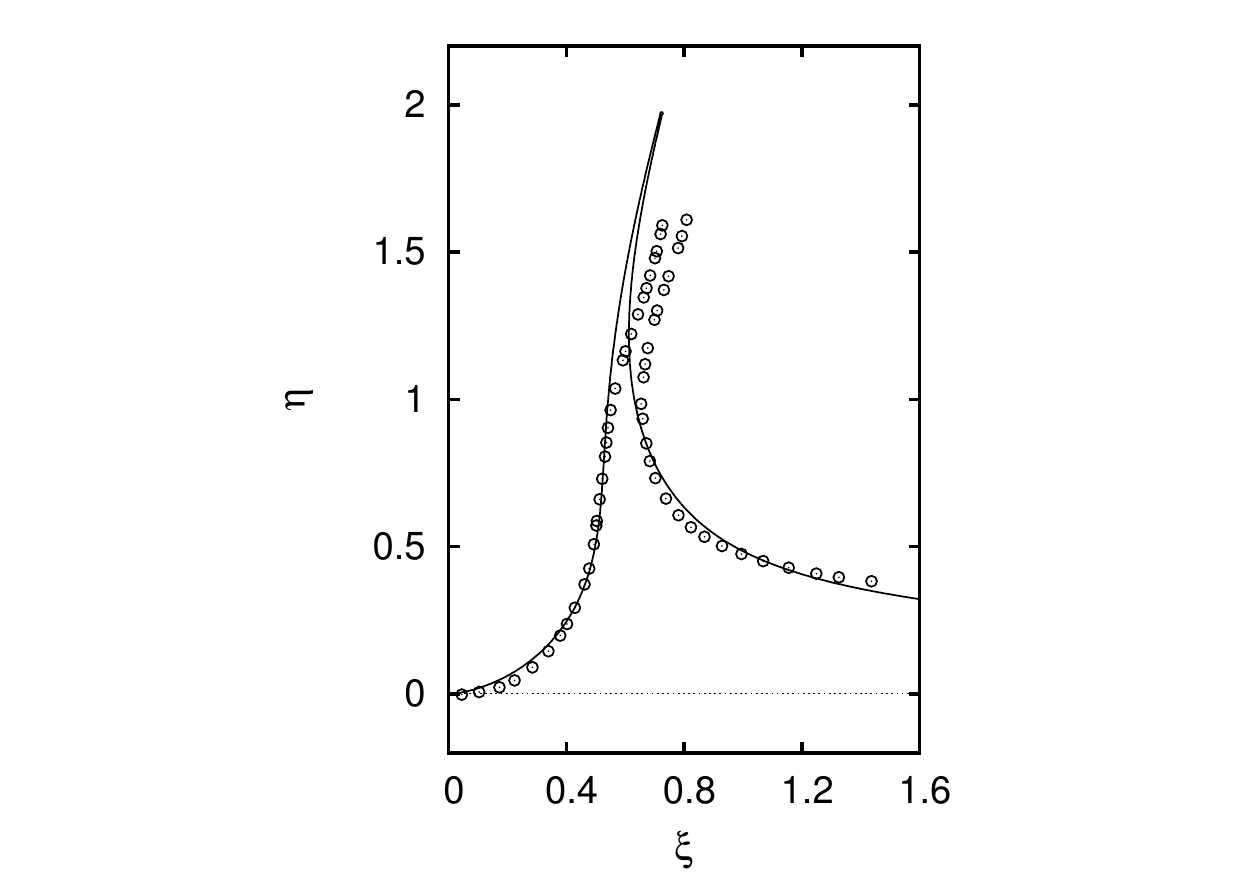}
}}
\caption{On the left, the convergence history of the free surface shape
nearby the plate edge is shown. On right, the final solution is compared
to the experimental data by \cite{yakimov}.\label{cfg_plate}}
\end{figure}

In Fig.~\ref{cfg_plate} the convergence history of the free surface
profiles is shown, along with a comparison of the final free surface
shape with experimental data \citep{yakimov}. In order to make more
evident the differences in the curves, a different scale is used for the
horizontal and experimental axes in the picture with the convergence
history. In establishing the comparison with the experimental
data, the points digitalized by the original paper are assumed with origin 
at the plate edge and are scaled by the same factor in both
directions, with the scale factor chosen to reach the best overlapping
at the root of the spray.

Once the convergence is achieved, the distribution of the velocity
potential along the body can be used to derive the pressure on the plate.
Starting from the definition of the stretched variables, the pressure
can be defined as $ p/\varrho = 1/B t^{-2/3} \psi$, where
\begin{equation}
\label{preself}
\psi = -\varphi + 2 (\xi \varphi_\xi + \eta \varphi_\eta) -
(\varphi_\xi^2 + \varphi_\eta^2) \;\;,
\end{equation}
the first two contributions originating from the time derivative and 
the third one related to the squared velocity term. It is worth noticing the
differences in the coefficients with respect to those found in the 
effective self-similar problem (\ref{pressure}).

\begin{figure}
\centerline{\hbox{
\includegraphics[width=80mm]{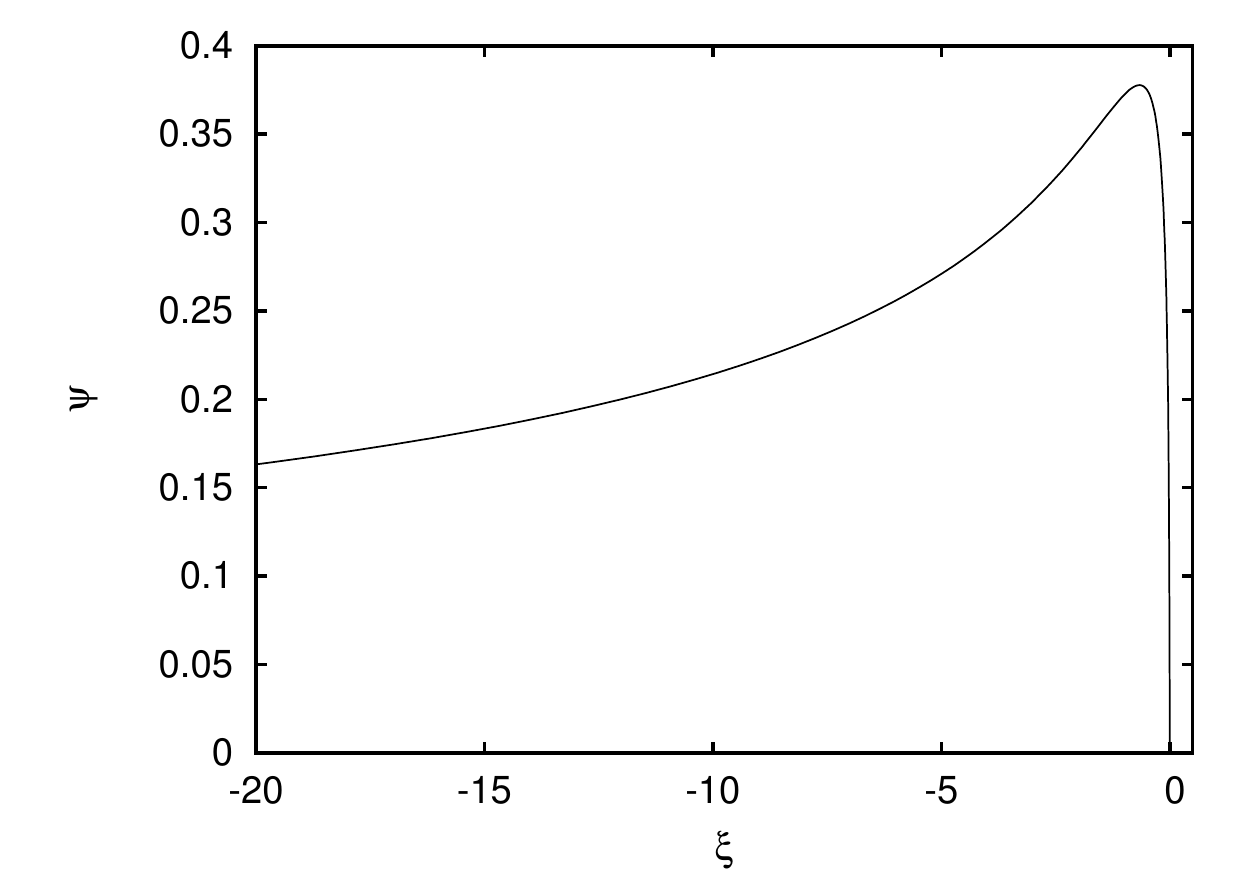}
}}
\caption{Pressure distribution acting on the plate.  \label{press_plate}}
\end{figure}

Whereas a more detailed discussion on the behaviour of the pressure and
the matching between inner and outer solutions is provided in \cite{iafrati08,
iafrati11}, here the attention is focussed on the pressure $\psi$ given in
terms of the inner variables.
The pressure, which is shown in Fig.~\ref{press_plate}, displays a peak 
located at a short distance from the edge and a sharp drop to
zero at the plate edge. On the other side, the pressure gently
diminishes approaching the outer solution.

\subsection{Ditching plate}

As a last example, the solution method is applied to derive the self-similar 
solution characterizing the water entry of a two-dimensional plate with
a high horizontal velocity component, which is considered as an 
exemplification of the aircraft ditching problem. 
There are two parameters governing the solution in this case, which are
the velocity ratio $V/U$ and the angle $\gamma$ formed by the plate with the
still water level.

The problem is self-similar under following set of variables
\begin{equation}
\label{selfvardi}
\xi = \frac{x}{Ut} \qquad
\eta = \frac{y}{Ut} \qquad
\varphi = \frac{\phi}{U^2t} \;\;,
\end{equation}
where $U$ and $V$ are the horizontal and vertical velocity of the plate.
In terms of the variables (\ref{selfvardi}), the plate edge is located 
at $(1,V/U)$.
The governing equations are about the same as (\ref{lapss})-(\ref{sff}),
aside from the symmetry condition, which does not hold in the ditching
case.
The boundary condition on the body is
\begin{equation}
\label{bbcssdi}
\varphi_\nu = \sin \gamma + \frac{V}{U} \cos \gamma \qquad \eta
=-\frac{V}{U} + (\xi-1) \tan \gamma  \;\;.
\end{equation}
As for the plate entry problem, also in this case an additional
condition is enforced at the edge, requiring that the free surface is
always attached to the edge and that the free surface leave the plate
tangentially. The solution on the left hand side is then rather similar
to that of the plate entry case, with the dynamic boundary condition
changing the sign at some inversion point. The position of the inversion 
point is determined by enforcing the two additional conditions at the
plate edge.

\begin{figure}
\centerline{\hbox{
\includegraphics[width=120mm]{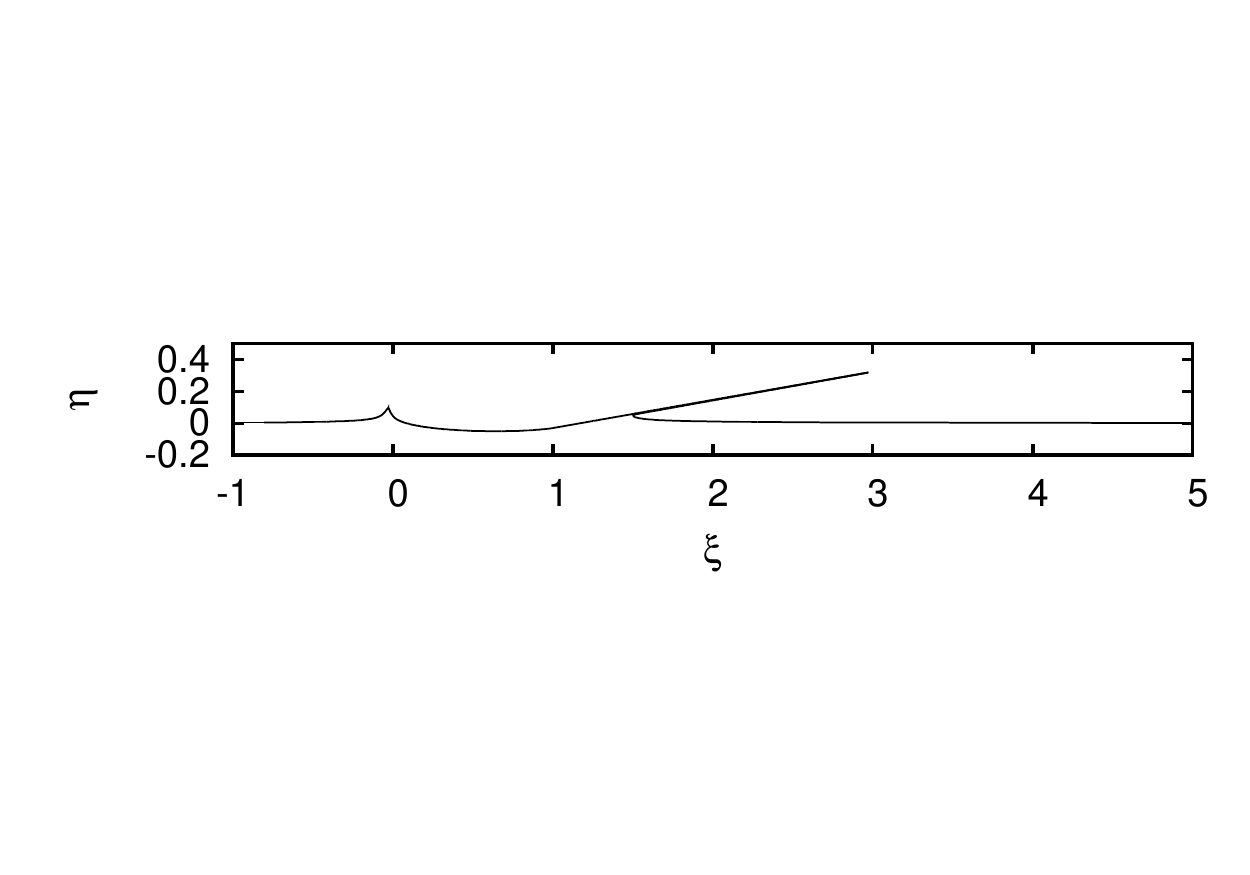}
}}
\centerline{\hbox{
\includegraphics[width=80mm]{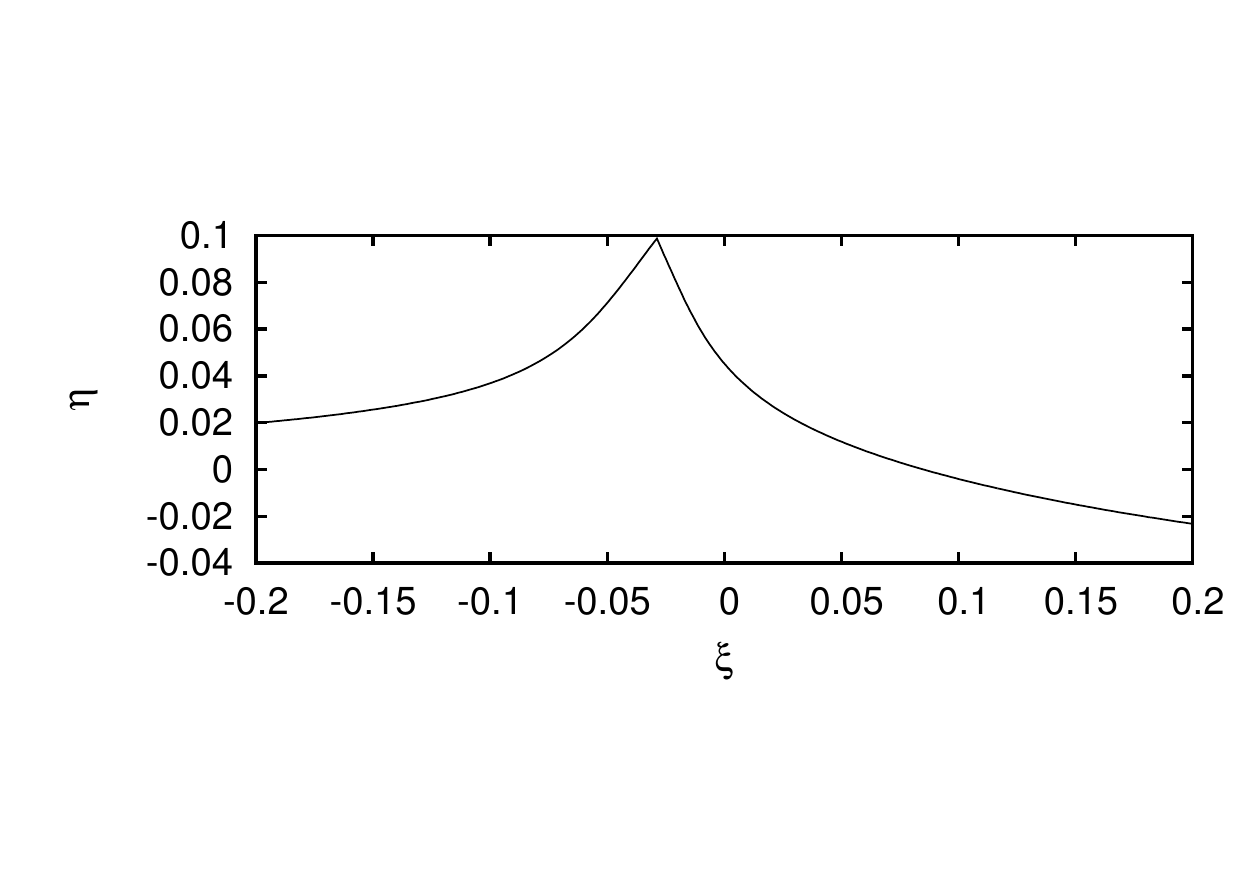}
}}
\caption{Free surface shape about a ditching plate, 10 degrees deadrise
angle. The plate is moving with a velocity ratio $V/U=0.03$.
Below, a close up view of the splash tip is provided.
\label{cfg_ditch}}
\end{figure}

In Fig.~\ref{cfg_ditch} the solution is shown in terms of free surface
shape for a plate with $\gamma=10$ degrees and a velocity ratio $V/U =0.03$.
The solution displays the very thin jet developing along the plate.
According to the definition of the stretched variables
(\ref{selfvardi}), the tip of the jet moves with a velocity which is
about three times the velocity of the plate. With similar
considerations, the root of the jet, which is just in front of the pressure 
peak, moves with a horizontal velocity which is about 1.5 times the
horizontal velocity of the plate.

Beside the thin jet developing along the plate, a splash is formed at
the rear. Differently from that found for the vertical entry case, the
splash is much milder and much thicker, with a rather large angle formed
by the two sides of the free surface at the tip.
For the case presented here, the tip of the splash
is located at $(-0.029,0.099)$, and the free surface is inclined 
of about 52 and 67 degrees with respect to the still water level on the
left and right hand side, respectively. This result in an internal angle
of about 61 degrees.

\begin{figure}
\centerline{\hbox{
\includegraphics[width=80mm]{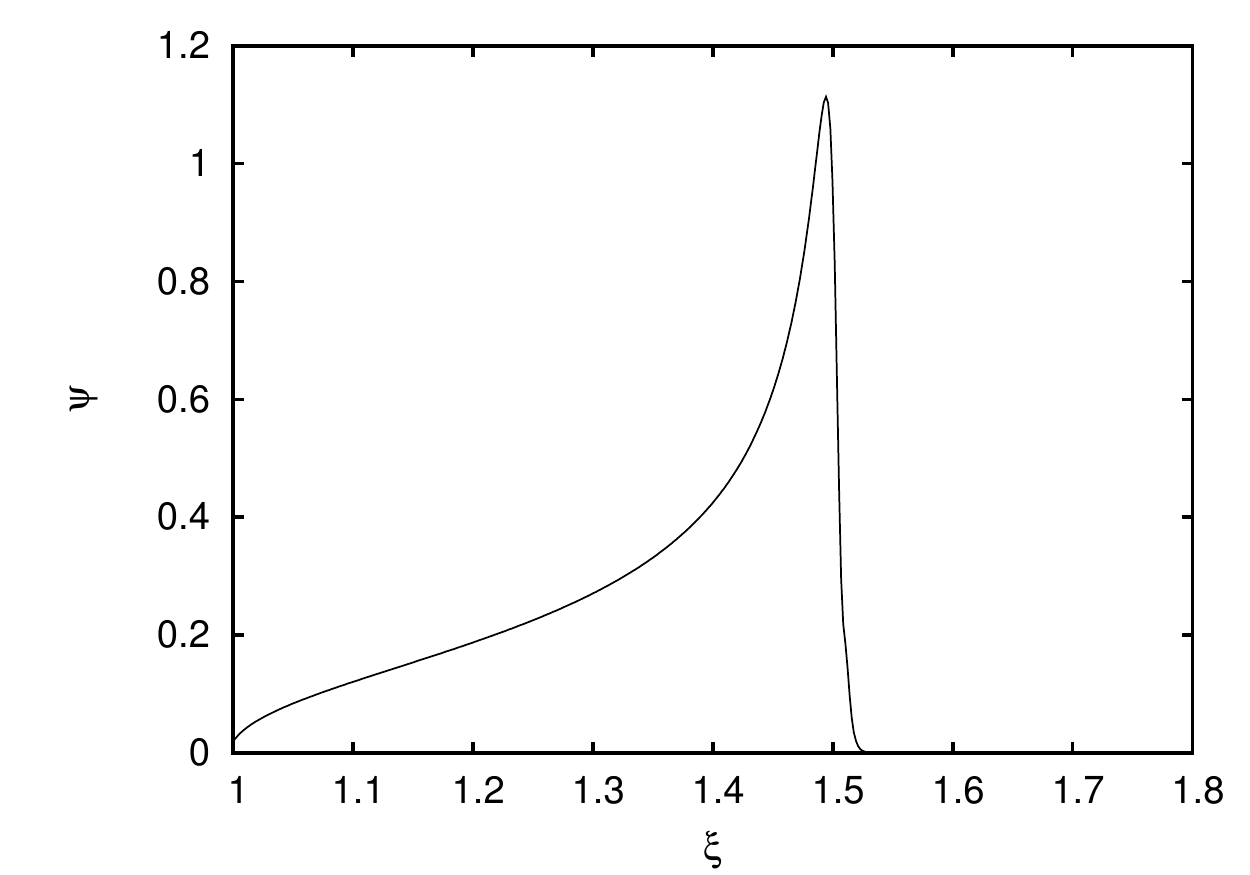}
}}
\caption{Pressure distribution about a ditching plate, 10 degrees
deadrise angle, with velocity ratio  $V/U=0.03$.  \label{press_ditch}}
\end{figure}

The pressure distribution for the same case is plotted in 
Fig.~\ref{press_ditch}, where $\psi = p/(\varrho U^2)$. 
The pressure peak is located at $\xi \simeq 1.49$, and thus the distance
to the plate edge is $0.49/\cos 10 \simeq 0.4976$. In the physical
variables this means that in a frame of reference attached to the body, 
the peak moves along the body surface with a velocity equal to $0.4976
U$.

Another important information that can be derived by the pressure
distribution is the total hydrodynamic load acting on the plate. 
By integrating the pressure distribution it is obtained that 
\[
F = \int p {\mbox d}s = (\varrho U^3 t) \int \psi {\mbox d}\tau
\]
where $\tau $ is the arclength measured along the body in the
self-similar plane. From the numerical integration of the pressure
distribution along the wetted surface it is obtained that $F \simeq
0.156 (\varrho U^3 t) $.

\section{Conclusions}

An iterative method for the solution of the fully nonlinear boundary value
problems characterizing self-similar free surface flows has been
presented.
The method has been applied to different examples characterized by
different boundary conditions in order to demonstrate the good level of
flexibility and accuracy. 
It has been shown that the method keeps a good accuracy even when dealing 
with the thin jets developing during water entry processes. In this
regard, the shallow water model proved to be very efficient, thus allowing 
a significant reduction of the computational effort without reducing the
level of accuracy.

The applications presented here are all referred to constant velocity. 
Future extension of the method may involve constant acceleration cases,
as that discussed in \cite{needham08}. In those cases however,
additional and more stringent hyphotheses are needed in order to
guarantee that gravity and surface tension effects are still
negligible.

\section*{Acknowledgement}

The author wish to thank Prof. Alexander Korobkin for the useful
discussions had during the development of the mathematical model and 
algorithm.
The part about the ditching problem has been done within the SMAES-FP7 project
(Grant Agreement N. 266172)

\bibliographystyle{model2-names}
\bibliography{<your-bib-database>}

%% not want to use model2-names.bst.


\clearpage

\clearpage

\clearpage

\clearpage

\end{document}